\def\lesssim{\mathrel{\hbox{\rlap{\hbox{\lower4pt\hbox{$\sim$}}}\hbox{$<$}}}}

\def\gtrsim{\mathrel{\hbox{\rlap{\hbox{\lower4pt\hbox{$\sim$}}}\hbox{$>$}}}}

\def\msun{\mbox{$M_{\odot}$}}

\def\ll_lsun{log$({L/\rm L_{\odot}})$~}

\def\masa_msun{$M/ \rm M_{\odot}$~}

\def\m_mstar{$M/M_{*}$~}

\documentclass{aa}
\usepackage{graphicx}

\begin{document}

\title{Asteroseismic inferences on 
GW Vir variable stars in the frame of new PG 1159 evolutionary models}

\author{A. H. C\'orsico$^{1,2}$\thanks{Member of the Carrera del Investigador
Cient\'{\i}fico y Tecnol\'ogico, CONICET, Argentina.} \and
L. G. Althaus$^{1,2 \star}$}

\offprints{A. H. C\'orsico}

\institute{$^1$Facultad  de  Ciencias Astron\'omicas  y Geof\'{\i}sicas, 
Universidad  Nacional de  La Plata,  Paseo del  Bosque s/n,  (1900) La
Plata, Argentina.\\ $^2$Instituto  de Astrof\'{\i}sica La Plata, IALP,
CONICET\\
\email{acorsico,althaus@fcaglp.unlp.edu.ar}}
\date{Received; accepted}

\abstract{An adiabatic, nonradial pulsation study  of GW Vir stars
is presented.   The pulsation calculations are based on PG1159 
evolutionary sequences with different stellar massess artificially 
derived from a full
evolutionary sequence of 0.5895 \msun~ that has been computed taking into 
account the evolutionary history of the progenitor star. The artificial 
sequences were constructed by appropriately  scaling
the stellar mass of the 0.5895-\msun sequence well  before the models
reach the low-luminosity, high-gravity stage of the GW Vir domain.
We compute  $g$-mode pulsation  periods  appropriate to  GW Vir  variable
stars.  The  implications for the  mode-trapping properties of  our PG
1159 models are  discussed at length. We found  that the mode-trapping
features characterizing  our PG  1159 models are  mostly fixed  by the
stepped shape  of the core  chemical profile left by  prior convective
overshooting.  This  is particularly  true at least  for the  range of
periods observed in  GW Vir stars. In addition,  we make asteroseismic
inferences  about  the internal  structure  of  the  GW Vir  stars  PG
1159-035, PG 2131+066, PG 1707+427 and PG 0122+200.
}

\authorrunning{C\'orsico \& Althaus}

\titlerunning{The GW Vir variable stars}

\maketitle


\section{Introduction}

GW Vir  (or DOV) stars constitute  currently one of  the most exciting
class  of  variable  stars,   since  they  represent  an  evolutionary
connection between  the cool,  very luminous, asymptotic  giant branch
(AGB) stars  and the  hot, extremely compact  white dwarf  (WD) stars.
Since  the discovery  of their  prototype, PG  1159-035, by  McGraw et
al. (1979), GW Vir stars have been the focus of numerous observational
and theoretical efforts.  These stars  are believed to be the variable
members of the PG 1159 family, a spectral class of stars characterized
by a  rather abnormal  surface chemical composition,  with atmospheres
devoid of hydrogen and, instead, rich in helium ($\sim 42
\%$),  carbon ($\sim  43 \%$)  and oxygen  ($\sim 15  \%$)  (Werner et
al.   1997).  PG 1159  stars ($80\,000  \lesssim T_{\rm  eff} \lesssim
180\,000$ K and $5.5 \lesssim \log  g \lesssim 8.0$) are thought to be
the  descendants of  post-AGB stars  that, after  experiencing  a late
thermal pulse at the beginning of  the WD cooling track, return to AGB
and finally evolve into the  hot central stars of planetary nebulae as
hydrogen-deficient  pre-WD  stars  ---  the  so-called  ``born-again''
scenario  (Fujimoto 1977; Sch\"onberner  1979, Iben  et al.  1983). In
fact,  recent  evolutionary calculations  of  the born-again  scenario
incorporating   convective  overshooting   have  been   successful  in
reproducing  the observed  photospheric composition  of PG  1159 stars
(Herwig et al. 1999; Althaus et al. 2005).

\begin{figure*}
\centering
\includegraphics[clip,width=400pt]{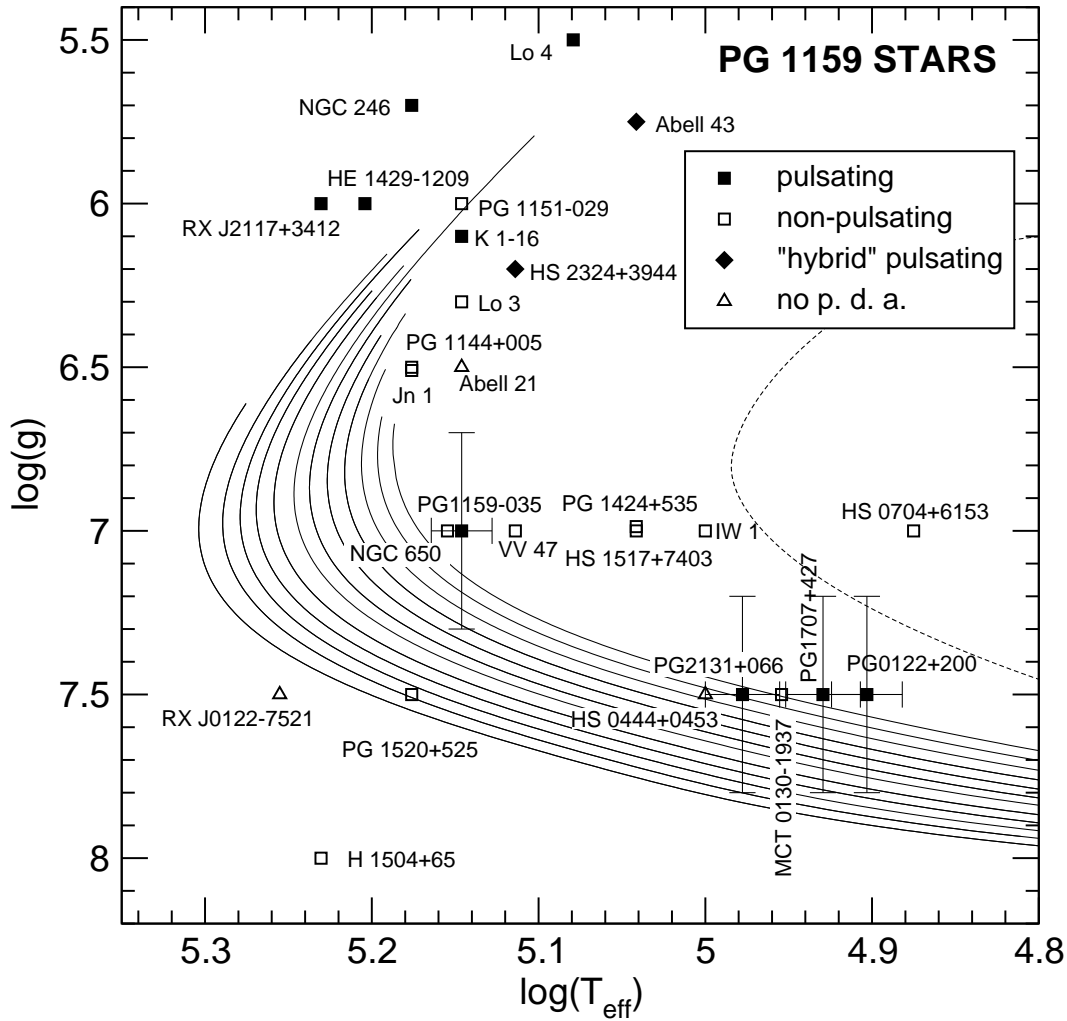}
\caption{The observed location of  PG 1159 stars 
in the  $\log T_{\rm eff}-\log  g$ plane. Filled squares  represent GW
Vir pulsators,  hollow squares represent non-pulsators  PG 1159 stars,
diamonds  correspond   to  ``hybrid-PG  1159''   stars  and  triangles
represent  stars with no  photometric data  available. Data  are taken
from Kawaler  et al. (2004), Werner  et al. (1997),  Dreizler \& Heber
(1998), Nagel  \& Werner (2004), and  Vauclair et al.  (2005). PG 1159
evolutionary tracks --- this work;  see also Althaus et al. (2005) ---
are indicated with solid curves; stellar masses are from right to left
$0.53, 0.54, \ldots, 0.64  M_{\odot}$.  Dashed curve corresponds to an
AGB-manqu\`e  evolutionary  track of  a  model  star  with $M_*=  0.49
M_{\odot}$ (Sc\'occola et al. 2005).  In this and the remainig figures
displaying  results from  our additional  evolutionary  sequences with
different stellar  masses, only results for  evolutionary stages after
the transitory phase has long  vanished are shown (see text toward the
end of \S \ref{2}).}
\label{grav-teff}
\end{figure*}

GW Vir variables  are low-amplitude, multiperiodic $g$-mode pulsators,
with  periods in  the range  from  5 to  30 minutes.   Eleven of  such
variables  are  presently  known,  six  of  which  associated  with  a
planetary nebula  and sometimes  termed PNNV (Planetary  Nebula Nuclei
Variable). The  remainder objects, lacking a  surrounding nebulae, are
called ``naked''  GW Vir  stars. In Fig.  \ref{grav-teff} we  plot the
observed  location of  the  known GW  Vir  stars in  the $\log  T_{\rm
eff}-\log g$ plane.  In addition,  we have included in the figure some
non-variable  PG  1159  stars as  well  as  PG  1159 stars  for  which
photometric  data  are not  available.  Note  that  GW Vir  stars  are
separated   into   two    subgroups,   one   containing   low-gravity,
high-luminosity stars ---  the PNNVs RXJ2117+3412, HE1429-1209, K1-16,
NGC  246,  HS2324+3944,  Lo  4   and  Abell  43  ---,  and  the  other
corresponding  to high-gravity  stars ---  the naked  GW Vir  stars PG
1159-035, PG 2131+066, PG 1707+427  and PG 0122+200.  Finally, we note
that among the low-gravity GW  Vir stars there are two objects labeled
as pulsating  ``hybrid-PG 1159'' ---  HS2324+3944 and Abell  43.  They
are variable PG 1159 stars  showing strong Balmer lines of hydrogen in
their spectra.

A longstanding problem associated with  GW Vir stars is related to the
excitation  mechanism for the  pulsations and,  in particular,  to the
chemical composition  of the  driving zone, an  issue that  has always
attracted the attention of the  theorists but that until very recently
has  eluded all the  attempts of  understanding.  In  their pioneering
works,  Starrfield et  al.   (1983, 1984,  1985)  and Stanghellini  et
al. (1991) realized that $g$-mode pulsations with periods around 500 s
could be driven in PG 1159 models by the $\kappa$-mechanism associated
with the partial ionization of  carbon and oxygen. While these authors
were successful  in finding  the correct destabilizing  mechanism, the
periods  of the  excited modes  were too  short as  compared  with the
observed  periodicities in  GW Vir  stars. In  addition,  their models
required a driving  region very poor in helium in  order to be capable
to  excite pulsations.  The  latter requirement  led to  the suspicion
that  a  composition  gradient  would  exist to  make  compatible  the
helium-devoid envelope  and the helium-rich  atmospheric layers.  Even
very recent theoretical work by  Bradley \& Dziembowski (1996) and Cox
(2003) points  out the  necessity  of a  chemical  composition in  the
driving region different from  the observed surface composition.  This
is difficult  to explain in  view of the  fact that PG 1159  stars are
still  experiencing mass  loss, a  fact  that prevents  the action  of
gravitational  settling of carbon  and oxygen.   Clearly at  odds with
conclusions  of  these works,  the  calculations  by  Saio (1996)  and
Gautschy  (1997), and  more recently  by  Quirion et  al.  (2004)  and
Gautschy et  al.  (2005), demonstrate that $g$-mode  pulsations in the
range of  periods of GW  Vir stars could  be easily driven in  PG 1159
models at satisfactory effective temperatures with an uniform envelope
composition compatible with the observed photospheric abundances.

While non-adiabatic stability issues like those mentioned above are of
utmost importance in elucidating the cause of GW Vir pulsations and to
place constraints  on the  composition of the  subphotospheric layers,
there is  other avenue that  may help to extract  valuable information
about the internal structure and evolutionary status of PG 1159 stars,
by  considering   {\it  adiabatic}  pulsation   periods  alone.   This
approach, known  as asteroseismology,  allows us to  derive structural
parameters of individuals pulsators by matching adiabatic periods with
the observed periodicities.  Current examples of asteroseismic studies
of GW Vir stars are those  of Kawaler \& Bradley (1994) (PG 1159-035),
Kawaler  et al.  (1995)  (PG  2131+066), O'Brien  et  al.  (1998)  (PG
0122+200),  Vauclair et al.  (2002) (RXJ2117+341)  and Kawaler  et al.
(2004) (PG 1707+427).  These studies have been successful in providing
precise  mass determinations  and  valuable constraints  on the  outer
compositional stratification  of PG 1159  stars. As important  as they
were, these investigations were based  on stellar models that were not
actually  fully  self-consistent   evolutionary  PG  1159  models.  In
addition, with the notable exception  of the exemplary work of Kawaler
\& Bradley  (1994), all  these studies used  the observed  mean period
spacing and the asymptotic theory  of stellar pulsations alone to make
asteroseismic  inferences; that  is, no  detailed period  fitting were
carried out in  those works. In part, the reason for  this is that for
most of GW Vir stars (with exception of PG 1159-035) the observed mode
density  appear  to be  not  high  enough  for detailed  asteroseismic
analysis.

It is  important to mention  that for applications  requiring accurate
values of  the adiabatic pulsation periods  of GW Vir  stars only {\it
full evolutionary} models can be  used.  This is particularly true for
asteroseismic fits to individual stars. In fact, as it has been shown,
stellar models  representing PG 1159 stars should  reflect the thermal
structure  of  their  progenitors  because  is  that  structure  which
primarily determines  the adiabatic period  spectrum, at least  at the
high  luminosity phases  (Kawaler et  al.  1985).   This is  of utmost
importance  concerning   the  theoretically  expected   rates  of  the
pulsation period  change ---  see, e.g., Kawaler  et al.   (1986).  In
addition, special care  must be taken in the  numerical simulations of
the  evolutionary stages prior  to the  AGB phase.   Specifically, the
modeling process should include extra mixing episodes beyond the fully
convective core during central helium burning (Straniero et al. 2003).
In particular, overshoot during the central helium burning stage leads
to sharp variations of the chemical  profile at the inner core --- see
Straniero et al. (2003); Althaus et al.  (2003).  This, in turn, could
have a  strong impact on the  mode-trapping properties of  the PG 1159
models.   Last  but not  least,  a  self-consistent  treatment of  the
post-AGB  evolution is  required. In  fact,  the details  of the  last
helium thermal pulse on the early WD cooling branch and the subsequent
born-again  episode  determines,  to  a  great  extent,  the  chemical
stratification and composition of the  outer layers of the hot PG 1159
stars.   The composition  stratification of  the stellar  envelope, in
turn, plays an important  role in the mode-trapping characteristics of
PG 1159 stars --- see Kawaler \& Bradley (1994).

All these requirements are completely fulfilled by the
0.5895-\msun~  PG 1159  models presented  recently by  Althaus  et al.
(2005).  In  fact,  these  authors  have  computed  the  evolution  of
hydrogen-deficient  post-AGB WDs  taking into  account a  complete and
detailed  treatment  of  the  physical  processes  that  lead  to  the
formation  of such  stars. These  models have  been recently  used for
non-adiabatic studies of  GW Vir stars by Gautschy  et al. (2005). The
aim  of  this  paper  is  to explore  the  {\it  adiabatic}  pulsation
properties of GW Vir stars  by employing the full evolutionary PG 1159
models of Althaus et  al. (2005). In  addition, we  employ these
models to make seismic inferences  on several GW Vir stars.  The paper
is  organized  as  follows.   In   Sect.   2,  we  describe  the  main
characteristics  of  our  numerical  codes  and  the  PG  1159  models
employed.   The  adiabatic  pulsational  properties for  our  PG  1159
sequences are described in detail  in \S 3.  Special emphasis is given
to the mode-trapping properties of  our models.  Sect. 4 is devoted to
describing the  application of  our extensive period  computations for
asteroseismic inferences of the  main structural properties of various
high-gravity GW Vir stars.  Finally,  in \S 5 we briefly summarize our
main findings and draw our conclusions.

\begin{figure}
\centering
\includegraphics[clip,width=250pt]{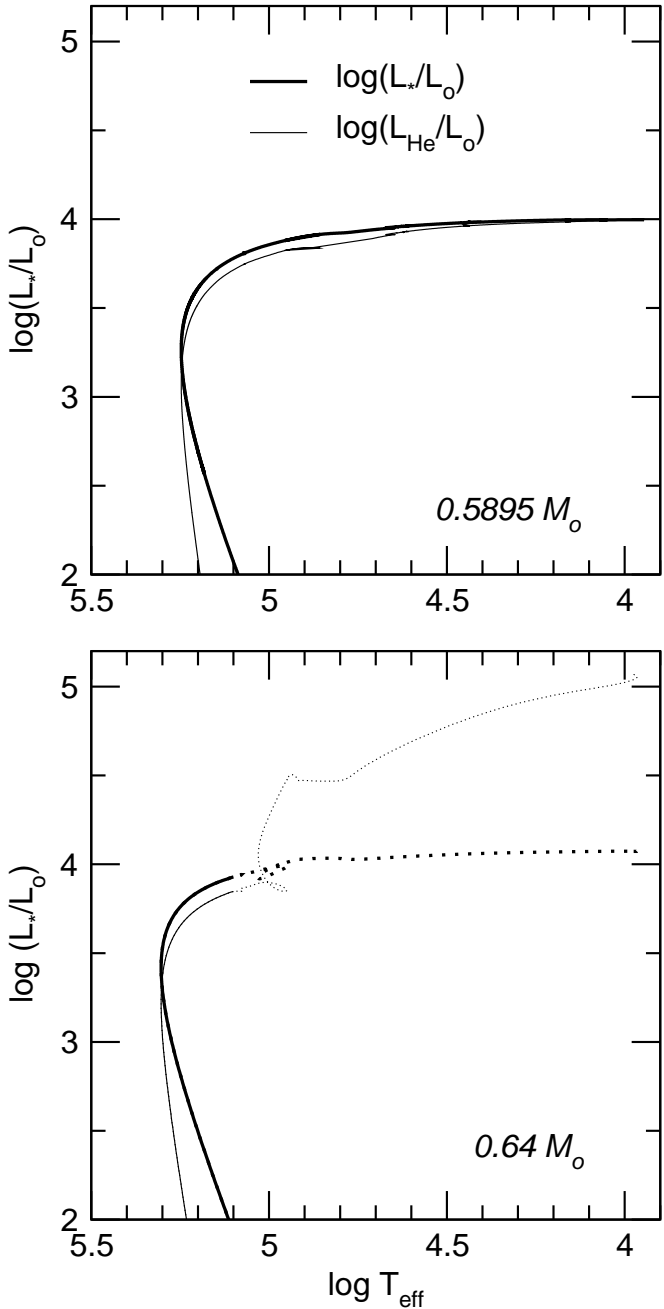}
\caption{The surface (thick line) and helium-burning (thin line) 
luminosities ($\log L_*/L_{\odot}$ and $\log L_{\rm He}/L_{\odot}$, 
respectively) in terms of the effective 
temperature. The upper panel depicts the situation for our consistent 
0.5895-\msun~  sequence. The bottom panel corresponds to the 0.64-\msun~  
artificially generated sequence, in which the the unphysical 
evolutionary stages are indicated with dotted lines. 
For details, see text.}
\label{demonstration}
\end{figure}

\section{Evolutionary code, input physics and 
stellar models for GW Vir stars}
\label{2}

The adiabatic pulsational analysis presented  in this work is based 
on artificial PG1159 evolutionary sequences with different 
stellar massess.  These sequences have been derived 
from a full post-born again evolutionary sequence 
of 0.5895 \msun~ (Althaus et  al.  2005) that has been computed taking into 
account the evolutionary history of the progenitor star. 
Specifically, the evolution  of an initially $2.7\, M_{\sun}$
stellar model  from the zero-age  main sequence has been  followed all
the way from the stages of  hydrogen and helium burning in the core up
to  the tip  of  the AGB  where  helium thermal  pulses occur. 
We mention that during the thermally pulsing AGB phase our models 
experience appreciable third dredge-up which causes the 
carbon-oxygen ratio to increase from $\approx 0.25$ to 
$\approx 0.34$ by the time the remnant departs from the AGB. 
Apart from implications for the chemical stratification of the 
post-AGB remnant, the 
occurrence of the third dredge-up mixing event affects the degeneracy 
of the carbon-oxygen core, which in turn affects the location of the 
PG1159 tracks (see Werner \& Herwig 2006 for comments).  
After experiencing 10  thermal pulses, the  progenitor departs from  the AGB
and evolves toward high effective temperatures, where it experiences a
final thermal pulse during the early  WD cooling phase --- a very late
thermal pulse and the ensuing born-again episode, see Bl\"ocker (2001)
for  a review  --- where  most of  the residual  hydrogen  envelope is
burnt.    After    the   occurrence   of   a    double-loop   in   the
Hertzsprung-Russell  diagram,  the  now hydrogen-deficient,  quiescent
helium-burning  0.5895-\msun   remnant  evolves  at   almost  constant
luminosity  to the domain  of PG  1159 stars  with a  surface chemical
composition     rich     in     helium,     carbon     and     oxygen:
($^{4}$He,$^{12}$C,$^{16}$O)=    (0.306,   0.376,   0.228)\footnote{In
particular,  the  surface abundance  of  oxygen  depends  on the  free
parameter $f$ of overshooting (which is a measure of the extent of the
overshoot region)  and the number of thermal  pulses considered during
the  AGB phase (see  Herwig 2000).}.  This is  in good  agreement with
surface  abundance  patterns  observed  in  pulsating  PG  1159  stars
(Dreizler  \& Heber 1998;  Werner 2001).   Also, the  surface nitrogen
abundance (about 0.01 by mass) predicted by our models is in line with
that detected in pulsating PG 1159 stars (see Dreizler \& Heber 1998).

We  mention  that  the  nuclear  network  considered  in  the  stellar
modelling  accounts  explicitly  for   16  chemical  elements  and  34
thermonuclear reaction rates to  describe hydrogen and helium burning.
Abundances changes  for the elements considered was  included by means
of  a   time-dependent  scheme  that   simultaneously  treats  nuclear
evolution and mixing processes  due to convection and overshooting.  A
treatment of this kind  is particularly necessary during the extremely
short-lived phase of the  born-again episode, for which the assumption
of instantaneous mixing is  inadequate. In particular, overshooting is
treated as a diffusion process (see, e.g., Herwig et al. 1997) and has
been considered  during all evolutionary  phases.  Radiative opacities
are  those of  OPAL (including  carbon- and  oxygen-rich compositions;
Iglesias \& Rogers 1996),  complemented, at low temperatures, with the
molecular opacities from Alexander \& Ferguson (1994).

\begin{figure}
\centering
\includegraphics[clip,width=250pt]{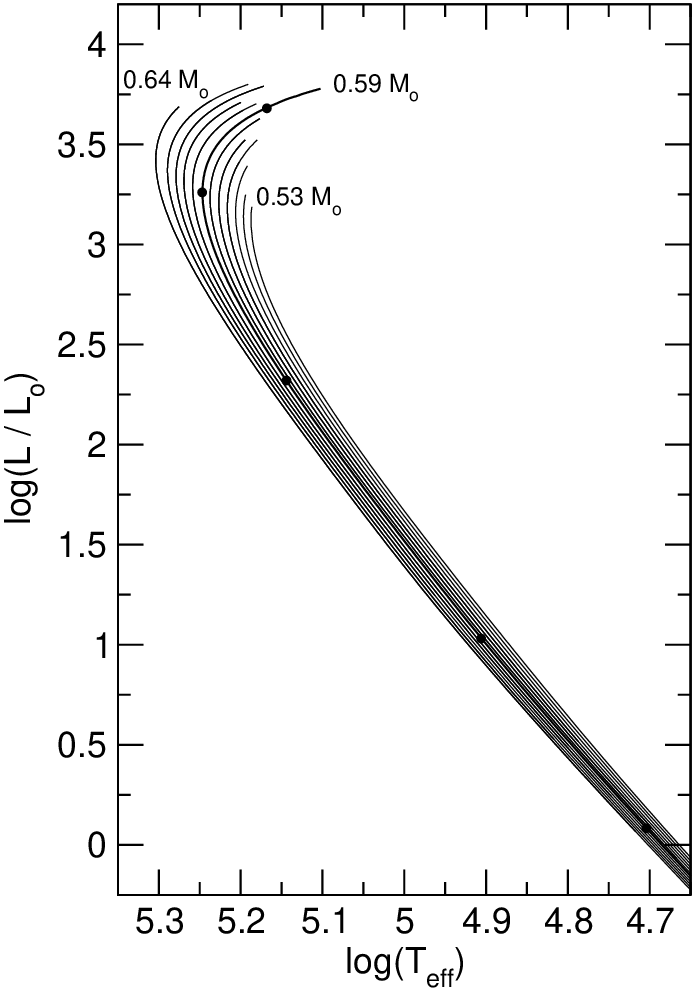}
\caption{Evolutionary tracks for sequences with stellar 
masses, from right  to left, of $0.53, 0.54,  \ldots, 0.64 M_{\odot}$.
The sequence with $0.5895 M_{\odot}$ is indicated with thicker line.
For this sequence, the black dots mark the location of five specific 
models to be analyzed in \S \ref{basic}.}
\label{tracks}
\end{figure}

Because  of the  considerable  load of  computing  time and  numerical
difficulty issues involved in the generation of PG 1159 stellar models
that take into account the  {\it complete} evolutionary history of the
progenitor star, we were forced  to consider one case for the complete
evolution, that  is, the 0.5895-\msun~ sequence.  However,  to reach a
deeper understanding of the  various trends exhibited by pulsations in
GW  Vir  stars, an  assessment  of  the  dependence of  the  pulsation
properties  of PG1159  models with  the several  structural parameters
should  be  highly desirable.   This  statement  is particularly  true
regarding the stellar mass of the models.  In this paper we consider a
set of additional sequences of  models within a small range of stellar
masses. In absence of actual self-consistent evolutionary computations
of  post-born again PG1159  models with  different stellar  masses, we
elect  to  create  several  evolutionary  sequences  by  employing  an
artificial  procedure starting  from  our ``seed''  model sequence  of
0.5895 \msun (see below).  This approach has been              adopted
in many pulsation studies aimed  at exploring the effects of different
model parameters  on the pulsation  properties (see, e.g.,  Kawaler \&
Bradley 1994), starting from  a {\it post-AGB} evolutionary model.  At
variance  with   previous  calculations,  in  this   work  we  derived
additional  model   sequences  with  stellar  masses   in  the  ranges
$0.53-0.58$ \msun~ and $0.60-0.64$ \msun~ (with a step of 0.01 \msun~)
from  the  {\it  post-born  again} 0.5895-\msun~  sequence  previously
described.   To   this  end,  we  artificially   changed  the  stellar
mass\footnote{We change the stellar mass in the following way.  In our
evolutionary  code,   the  independent  variable  is   $\xi=  \ln(1  -
m_r/M_*)$,  where $M_*$  is the  stellar mass  and $m_r$  is  the mass
contained in a sphere of radius  $r$. When the stellar mass is changed
(say to $M'_*$),  the $\xi$ values at each grid point  are the same as
before; so,  the new mass at  a given $\xi$  is $m'_r= M'_* (1  - \exp
(\xi))$. The chemical composition at each $\xi$ is the same as before,
but the mass contained at $\xi$ is different. For instance, if $M'_* <
M_*$, then $m'_r < m_r$ at a given $\xi$.  Note that this procedure is
different from,  for instance, simply  extracting mass from  the outer
layers.}   to the  appropriate values  shortly  after the  end of  the
born-again  episode.  Although  this procedure  led to  a  sequence of
unphysical stellar  models for which the  helium-burning luminosity is
not consistent with the  thermo-mechanical structure of the models, we
found that  the transitory stage  vanishes  at  high luminosities,
before the star reaches the  ``knee'' in the $\log T_{\rm eff}-\log g$
plane\footnote{This  procedure  has been  employed,  for instance,  by
O'Brien  (2000) on  a 0.573-\msun  post-AGB evolutionary  model.}  
This can be understood by examining Fig. \ref{demonstration} which shows
the surface and helium-burning luminosities in terms of the effective 
temperature. The upper panel depicts the situation for our consistent 
0.5895-\msun~  sequence. The bottom panel corresponds to the 0.64-\msun~  
artificially generated sequence. Note that, as result of the change 
in the thermal structure of the outer layers induced by the 
abrupt change in the stellar mass (at low effective temperatures), 
the helium-burning luminosity becomes seriously affected. Note that once  
the transitory stage has disappeared, the helium-burning luminosity
follows a similar trend to that expected from the 
full evolutionary computation (upper panel). In the bottom panel, 
the unphysical evolutionary stages are indicated with dotted lines, 
while the solid lines correspond to the evolutionary stages which 
we consider valid for the purposes of this paper. 
Thus, the tracks depicted in Fig.  \ref{grav-teff} correspond to 
those stages of
the  evolutionary sequences  for which  the unphysical  transitory has
long  disappeared. We  stress that  pulsation results  presented along
this  work  will be  performed  on  stellar  models belonging  to  the
evolutionary  sequences  shown   in  Fig.  \ref{grav-teff}  (see  also
Fig. \ref{tracks}). We warn  again that the pulsational 
results presented here are based on stellar models artificially 
constructed, and that full evolutionary models with different stellar 
masses and complete evolutionary history are required to place 
such results on a solid background.

We stress that, since we do  not account for the complete evolution of
the  progenitor stars  (except  for the  0.5895-\msun~ sequence),  the
internal composition of  the models we use in  part of our pulsational
analysis may not  be completely realistic, a fact  that should be kept
in  mind  when  we  analyze,  in particular,  the  dependence  of  the
asymptotic  period  spacing  and  mode trapping  properties  with  the
stellar mass. We also  stress that the shape of the carbon/oxygen
profile ---  particularly in  the core ---  is quite sensitive  to the
adopted reaction  rates during the  central helium burning  phase (see
Kunz et  al. 2002; Straniero  et al. 2003)  and the efficiency  of the
convection-induced  mixing  (overshooting  and/or semiconvection;  see
Straniero  et  al.  2003).   In  particular, we  have  recently  found
(C\'orsico  \&   Althaus  2005)  that   with  a  lower  rate   of  the
$^{12}$C$(\alpha,\gamma)^{16}$O  reaction than  we  adopted here,  the
central  abundances  of  oxygen   and  carbon  become  quite  similar,
partially smoothing the chemical steps  in the core. However, as shown
in that paper, the mode  trapping structure resulting from the core is
not seriously altered.

\section{Adiabatic pulsational analysis}

In this section we present an extensive adiabatic analysis of our grid
of PG 1159 evolutionary models.   Since the most detailed and thorough
reference in the  literature on adiabatic pulsations of  PG 1159 stars
is the work of Kawaler  \& Bradley (1994) [hereinafter KB94], we shall
invoke  repeatedly that paper  throughout this  work to  compare their
results  with  our  own  findings.   We have  assessed  the  pulsation
properties of  about 3000  stellar models, the  stellar mass  of which
range  from  0.53  $M_{\odot}$  to  0.64  $M_{\odot}$,  embracing  the
well-established  mean  mass for  WDs.   Fig.  \ref{tracks} shows  the
evolutionary  tracks  of our  model  sequences.  All  of our  PG  1159
sequences  are characterized by  a helium  layer thickness  of $0.0052
M_{\odot}$ ---  as predicted by our evolutionary  calculations for the
0.5895-$M_{\odot}$  models (Althaus  et al.  2005). For  each analyzed
model  we have  computed $g$-mode  periods in  the range  $50-2500$ s,
although in some cases we have  extended the range of periods and also
included  $p$-mode computations.  We limited  our calculations  to the
degrees $\ell= 1,2$ because the  periods observed in pulsating PG 1159
stars  have been positively  identified with  $\ell= 1,2$  (see, e.g.,
Winget et al  1991; Vauclair et al. 2002).  Since  one of the purposes
of  this work  is to  get values  of adiabatic  periods as  precise as
possible, we  have employed about $2700-3000$  mesh-points to describe
our background stellar models.

\subsection{Pulsation code and adiabatic quantities}
\label{code}

We begin this  section with a brief description  of the pulsation code
employed  and  the   relevant  pulsation  quantities.   For  computing
pulsation modes of our PG 1159 models we use an updated version of the
pulsational code  employed in C\'orsico et al.   (2001).  Briefly, the
code, which  is coupled to the  LPCODE evolutionary code,  is based on
the general Newton-Raphson technique  and solves the full fourth-order
set  of  equations  governing  linear,  adiabatic,  nonradial  stellar
pulsations  following  the  dimensionless formulation  of  Dziembowski
(1971).  The pulsation  code provides the dimensionless eigenfrequency
$\omega_{k}$  --- $k$  being  the radial  order  of the  mode ---  and
eigenfunctions  $y_1,\cdots, y_4$.   From these  basic  quantities the
code computes the pulsation periods ($\Pi_k$), the oscillation kinetic
energy  ($K_k$),  rotation  splitting coefficients  ($C_{k}$),  weight
functions ($W_k$), and variational  periods ($\Pi_k^{\rm v}$) for each
computed   eigenmode.  Usually,   the   relative  difference   between
$\Pi_k^{\rm  v}$ and  $\Pi_k$ is  lower than  $\approx  10^{-4}$.  The
pulsation  equations,  boundary   conditions  and  relevant  adiabatic
quantities are given in Appendix A.

The   prescription   we   follow    to   assess   the   run   of   the
Brunt-V\"ais\"al\"a   frequency  ($N$)   is  the   so-called  ``Ledoux
Modified'' treatment --- see  Tassoul et al.  (1990) --- appropriately
generalized  to include the  effects of  having three  nuclear species
(oxygen, carbon and  helium) varying in abundance (see  KB94). In this
numerical  treatment  the  contribution  to  $N$ from  any  change  in
composition  is almost completely  contained in  the Ledoux  term $B$;
this fact renders the method particularly useful to infer the relative
weight that each chemical  transition region have on the mode-trapping
properties  of the model  (see \S  \ref{modetrapping}).  Specifically,
the Brunt-V\"ais\"al\"a frequency is computed as:

\begin{equation}
N^2 = \frac{g^2\ \rho}{p}\ \frac{\chi_{\rm T}}{\chi_{\rho}}\
\left(\nabla_{\rm ad} - \nabla + B \right)
\label{bruntva}
\end{equation}

\noindent where 

\begin{equation}
B=-\frac{1}{\chi_{\rm T}} \sum^{n-1}_{i=1} \chi_{{\rm X}_i}
\frac{d\ln X_i}{d\ln p}, 
\label{bledoux}
\end{equation}

\noindent being

\begin{equation}
\chi_{\rm T}= \left[\frac{\partial \ln{p}}
{\partial \ln{T}} \right]_{\rho}\ \
\chi_{\rm \rho}= \left[\frac{\partial \ln{p}} 
{\partial \ln{\rho}} \right]_T\ \
\chi_{{\rm X}_i}= \left[\frac{\partial \ln{p}}
{\partial \ln{X_i}} \right]_{\rho,T,{X_{j \neq i}}}
\end{equation}

\begin{figure}
\centering
\includegraphics[clip,width=250pt]{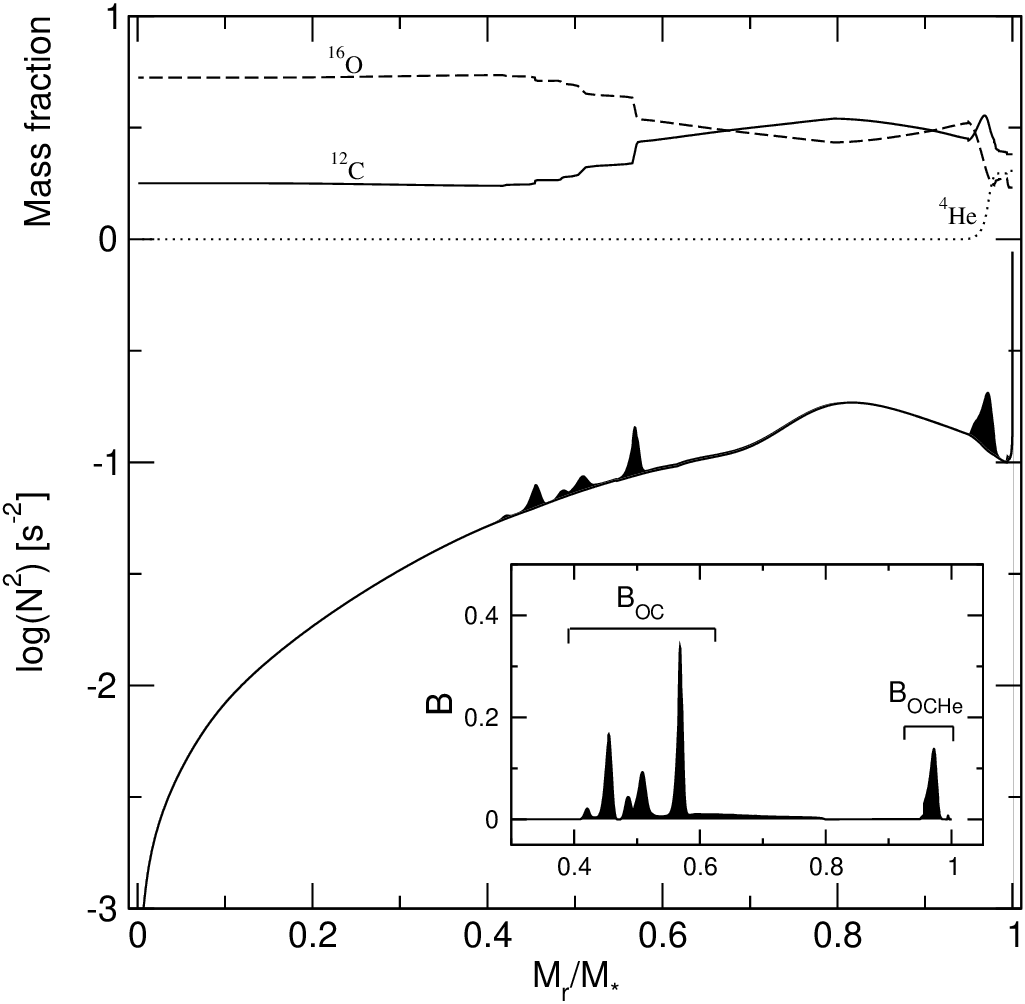}
\caption{The run of the squared Brunt-V\"ais\"al\"a frequency in terms 
of  the mass  coordinate, corresponding  to a  $0.5895$-\msun  PG 1159
model at  $T_{\rm eff}=  139\,000$ K and  $\log(L_*/L_{\odot})= 2.31$.
The  mass  of  the  helium  content  is  of  $0.008821  M_*$  ($0.0052
M_{\odot}$).  Dark regions denote the contributions of the Ledoux term
$B$ (shown  in the inset)  to the Brunt-V\"ais\"al\"a  frequency.  The
chemical profile is displayed in the upper zone of the plot.}
\label{profile}
\end{figure}

Fig.  \ref{profile}   shows  a  representative  spatial   run  of  the
Brunt-V\"ais\"al\"a  frequency for  a  PG 1159  model.   The model  is
characterized by a stellar mass of $0.5895$
\msun,  an effective  temperature of  $\approx 139\,000$  K and  a 
luminosity of $\log(L_*/L_{\odot})= 2.31$. In addition, the plot shows
the internal chemical stratification of the model for the main nuclear
species (upper region  of the plot) and for  illustrative purposes the
profile of  the Ledoux  term $B$ (inset).   The figure  emphasizes the
role   of   the   chemical    interfaces   on   the   shape   of   the
Brunt-V\"ais\"al\"a  frequency.   In  fact, each  chemical  transition
region  produces clear  and  distinctive features  in  $N$, which  are
eventually  responsible  for   the  mode-trapping  properties  of  the
model. At the core region  there are several peaks at $M_r/M_* \approx
0.4-0.6$ resulting  from steep  variations in the  inner oxygen/carbon
profile.   The  stepped  shape  of  the carbon  and  oxygen  abundance
distribution within the core is  typical for situations in which extra
mixing episodes beyond the fully convective core during central helium
burning are allowed. In particular, the sharp variation around $M_r
\approx 0.56 M_*$, which is left by  overshoot at the boundary of the 
convective core during  the central helium burning phase,  could be a
potential   source   of  mode-trapping   in   the   core  region   ---
``core-trapped'' modes; see  Althaus et al.  (2003) in  the context of
massive DA  WD models. The extended  bump in $N$ at  $M_r \approx 0.96
M_*$  is  other  possible   source  of  mode-trapping,  in  this  case
associated with  modes trapped  in the outer  layers. This  feature is
caused  by  the  chemical  transition  of helium,  carbon  and  oxygen
resulting from  nuclear processing in  prior AGB and  post-AGB stages.
In a subsequent Section we shall  explore in detail the role played by
the chemical transition regions on the mode-trapping properties of our
PG 1159 evolutionary models.

\subsection{The basic nature of the nonradial pulsation spectrum}
\label{basic}

This  section  is  devoted  to  exploring at  some  length  the  basic
properties  of  nonradial  pulsation  modes  of  our  PG  1159  pre-WD
evolutionary models.  To  this end, we have extended  the scope of our
pulsation  calculations  by including  low  order  $g$-modes and  also
nonradial  $p$-modes in  the discussion,  although we  are  aware that
short periods like those associated with such modes have not been ever
detected in variable  PG 1159 stars.  Here, we  have chosen to analyze
the  sequence of $0.5895  M_{\odot}$-models from  the early  phases of
evolution  at constant luminosity  --- shortly  before the  maximum of
effective temperature --- until the  beginning of the WD cooling track
--- when  models evolve at  roughly constant  radius.  We  stress that
this  is the  first time  that the  adiabatic pulsation  properties of
fully evolutionary post born-again PG  1159 models are assessed --- an
exception  is the  {\it  non-adiabatic} analysis  of  Gautschy et  al.
(2005).   Thus, we believe  that it  is instructive  to look  into the
basic  features of  the  full nonradial  pulsation  spectrum of  these
models.

\begin{figure}
\centering
\includegraphics[clip,width=250pt]{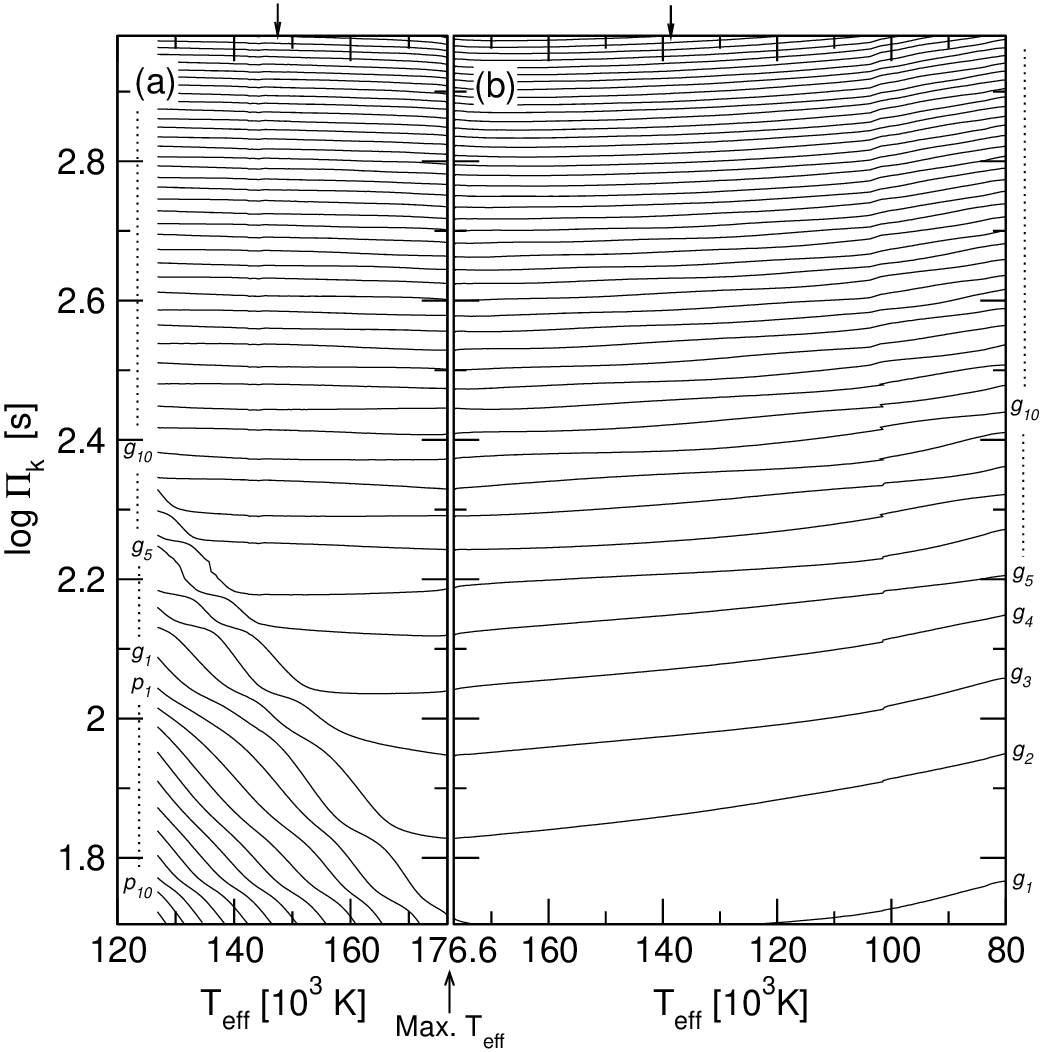}
\caption{The evolution of the $\ell= 1$ 
pulsation  periods  $\Pi_k$ in  terms  of  the effective  temperature,
corresponding to the sequence  of $0.5895 M_{\odot}$-models. Panel (a)
depicts the  situation before  the models reach  the maximum  value of
effective temperature  (Max. $T_{\rm eff}$),  and panel (b)  shows the
situation   when  the   models  have   already  passed   through  this
stage. Arrows indicate the  effective temperature corresponding to two
elected  models   that  are   analyzed  in  Figs.    \ref{propa1}  and
\ref{propa3}. Note that  the orientation of the $T_{\rm  eff}$ axis in
panel (a) is opposite to that of panel (b). See text for details.}
\label{per-teff}
\end{figure}

We begin by describing the  evolution of the period spectrum.  For the
$0.5895  M_{\odot}$ sequence we  have available  physically meaningful
stellar models from stages previous to the beginning of the WD cooling
track. This fact  allows us to perform an  exploratory exercise of the
period  spectrum from  the moment  when the  objects still  retain the
thermal  structure resulting  from prior  evolution of  the progenitor
star.  In Fig. \ref{per-teff} we  show the $\ell= 1$ pulsation periods
$\Pi_k$ in  terms of  the effective temperature  before (panel  a) and
after (panel b)  the model reach the turning point  in the H-R diagram
(see Fig. \ref{tracks}).  We find the pulsation periods to decrease in
stages  in  which  the  model  is still  approaching  to  its  maximum
effective  temperature ($T_{\rm  eff}  \approx 176\,600$  K). This  is
because the star  is undergoing a rapid contraction,  in particular at
the  outer layers. As  is well  known, contraction  effects lead  to a
decrease in the pulsation periods  (see Winget et al. 1983).  When the
model  has  already settled  into  their  cooling  phase, the  periods
increase  in  response  to  the decrease  of  the  Brunt-V\"ais\"al\"a
frequency.  Thus,  the behavior exhibited is typical  of WD pulsators,
with increasing  periods as the effective  temperature decreases. Note
in panel (a) of Fig. \ref{per-teff} that when $T_{\rm eff} < 176\,600$
K, the low  order periods ($\Pi_k \lesssim 200$  s) exhibit a behavior
reminiscent to  the well  known ``avoided crossing''.  When a  pair of
modes experiences avoided crossing, the modes exchange their intrinsic
properties (see Aizenman et al.   1977).  As we shall see below, these
are  modes with  a  mixed character,  that  is, modes  that behave  as
$g$-modes  in certain  zones of  the star  and as  $p$-modes  in other
regions.

In  order  to obtain  valuable  qualitative  information of  nonradial
pulsations we apply a local treatment to the pulsation equations (Unno
et  al.  1989).  Employing  the Cowling  approximation,  in which  the
perturbation of the gravitational potential is neglected ($\Phi'= 0$),
and  assuming that  the coefficients  of the  pulsation  equations are
almost constant --- that is, in the limit of high radial order $k$ ---
we obtain simple expressions for the eigenfunctions: $y_1, y_2 \propto
\exp (i k_r r)$, and an useful dispersion
relation follows: $k_r^2= \sigma^{-2} c^{-2} \left(\sigma^2 - N^2
\right)
\left( \sigma^2 -L_{\ell}^2\right)$, which relates the local radial wave 
number $k_r$  to the pulsation  frequency $\sigma$. $L_{\ell}$  is the
Lamb frequency,  defined as  $L_{\ell}^2= \ell (\ell  + 1)c^2  / r^2$.
Note  that  if  $\sigma^2  >  N^2, L_{\ell}^2$  or  $\sigma^2  <  N^2,
L_{\ell}^2$, the wave  number $k_r$ is real, and if  $N^2 > \sigma^2 >
L_{\ell}^2$ or $N^2 <
\sigma^2 < L_{\ell}^2$, $k_r$ is purely imaginary. 

\begin{figure}
\centering
\includegraphics[clip,width=250pt]{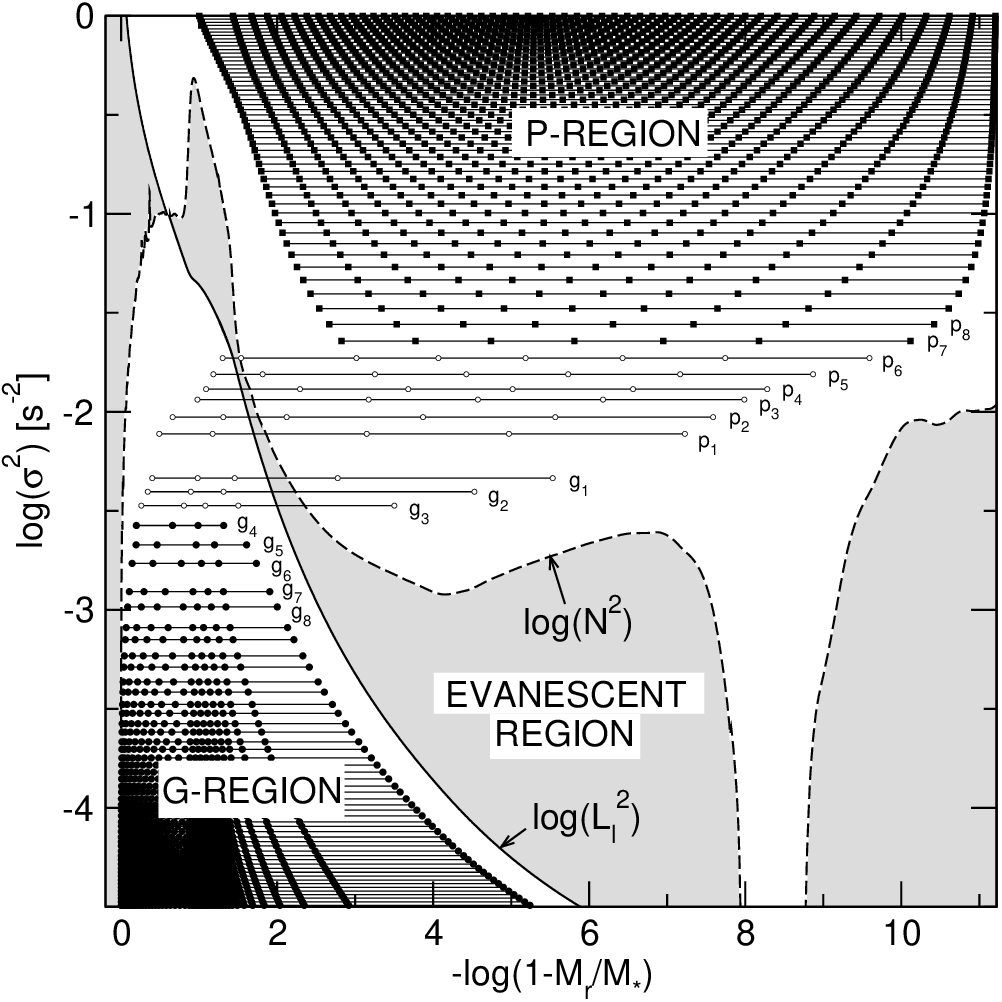}
\caption{Propagation diagram corresponding to a PG 1159 evolutionary 
model characterized by $[M_*/M_{\odot}, \log(L_*/L_{\odot}),
\log T_{\rm eff}]= [0.5895, 3.68, 5.168]$ for $\ell=  1$ modes.  
The effective  temperature of  this model is  marked with an  arrow in
panel (a)  of Fig.  \ref{per-teff}.  White (gray) areas  correspond to
propagation (evanescent) zones  according the local analysis described
in the  text.  The square  of the Brunt-V\"ais\"al\"a  frequency ($N$)
and the Lamb frequency ($L_{\ell}$) are depicted with dashed and solid
lines,  respectively.  Horizontal thin  lines show  the square  of the
eigenfrequencies, and small symbols show  the loci of the nodes of the
radial  eigenfunction,   $y_1$  (filled  circles:   $g$-modes;  filled
squares: $p$-modes).   Open circles correspond to nodes  of modes with
mixed  character.  These  modes propagate  in both  P-  and G-regions.
Note the  excellent agreement between the full  numerical solution and
the predictions of the simple local treatment.}
\label{propa1}
\end{figure}

\begin{figure}
\centering
\includegraphics[clip,width=250pt]{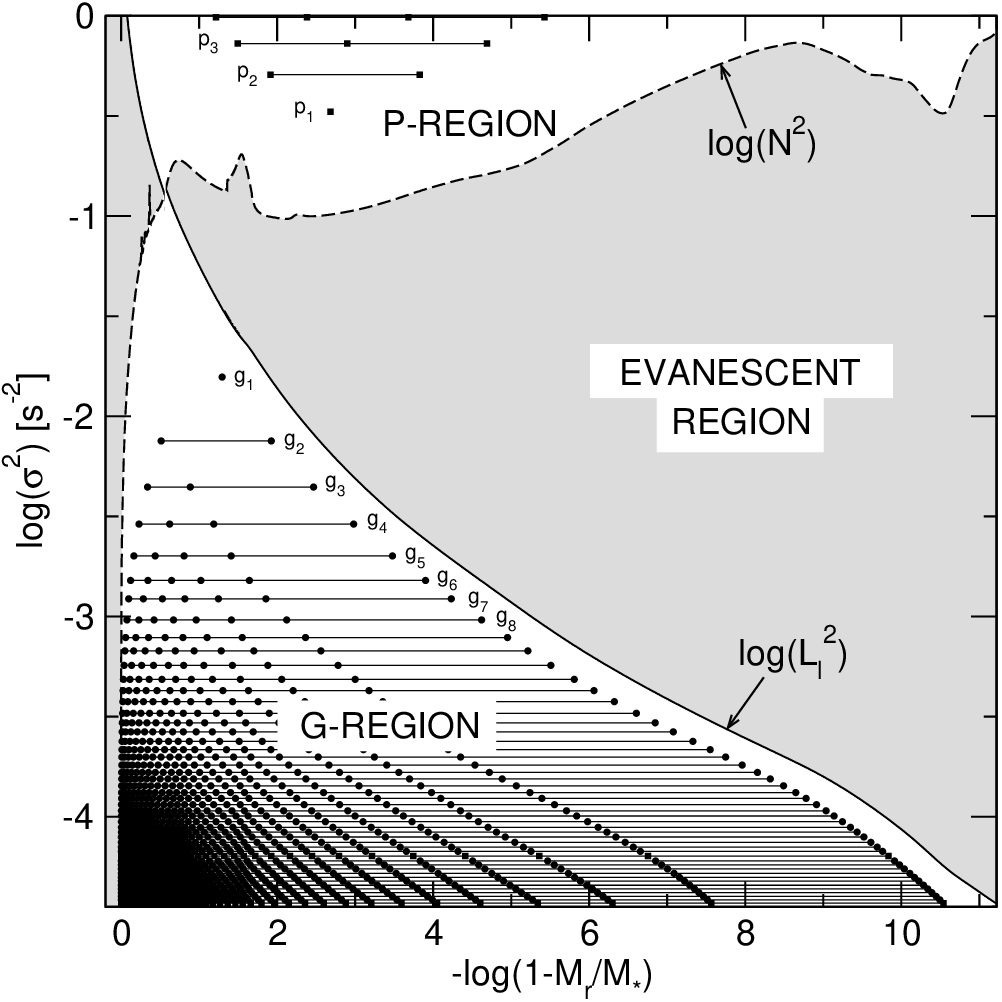}
\caption{Same of Fig. \ref{propa1}, but for a PG 1159 model 
characterized by $[M_*/M_{\odot}, \log(L_*/L_{\odot}),
\log T_{\rm eff}]= [0.5895, 2.31, 5.143]$. The effective temperature 
of  the  model  is  that  marked   with  an  arrow  in  panel  (b)  of
Fig. \ref{per-teff}.  Note that in this  case the G and  P branches of
modes are clearly separated in frequency.}
\label{propa3}
\end{figure}
                                                      
In Fig.  \ref{propa1}  we show $L_{\ell}^2$ and $N^2$  as functions of
$-\log  (1-M_r/m_*)$.    This  type   of  diagrams  are   called  {\it
propagation diagrams} (see,  e.g., Unno et al.  1989;  Cox 1980).  The
figure corresponds  to a PG  1159 model with $M_*=  0.5895 M_{\odot}$,
$\log(L_*/L_{\odot})= 3.68$ and $T_{\rm  eff}= 147\,200$ K. This model
corresponds to a  stage in which the star is  still approaching to its
maximum  $T_{\rm   eff}$  (see  the   arrow  in  panel  (a)   of  Fig.
\ref{per-teff}).    The  dashed   curve  shows   the  square   of  the
Brunt-V\"ais\"al\"a frequency, and the  solid curve depicts the run of
the  square of the  Lamb frequency  for $\ell=  1$.  Since  this model
still retains some similarities to  those of the red giants from which
these  models are  descendants, the  Brunt-V\"ais\"al\"a  frequency is
characterized   by  large   central  values   as  compared   to  those
characterizing the  outer regions. In the plot,  gray areas correspond
to regions  in which  $k_r$ is purely  imaginary.  As a  result, modes
should  be unable  to  propagate in  these  ``evanescent'' zones.   In
contrast, in regions in which $k_r$  is real (white areas in the plot)
modes should be oscillatory in  space, and these regions correspond to
``propagation'' zones.  As  we can see in the  figure, there exist two
propagation regions,  one corresponding to  the case $\sigma^2  > N^2,
L_{\ell}^2$, associated with $p$-modes  (P-region), and other in which
the   eigenfrequencies  satisfies   $\sigma^2   <  N^2,   L_{\ell}^2$,
associated with  $g$-modes (G-region). The predictions  of this simple
local analysis  are nicely confirmed  by our full  numerical solution.
Indeed,  the figure shows  the computed  eigenfrecuencies of  $p$- and
$g$-modes (thin  horizontal lines)  and the loci  of the nodes  of the
radial eigenfunction,  $y_1$ (small symbols).  Note that  no node lies
in the  evanescent regions, meaning  that the radial  eigenfunction is
not oscillatory  here.  This figure  clearly shows that  the $p$-modes
and  $g$-modes   are  trapped  in  the  P-region   and  the  G-region,
respectively.

\begin{figure}
\centering
\includegraphics[clip,width=250pt]{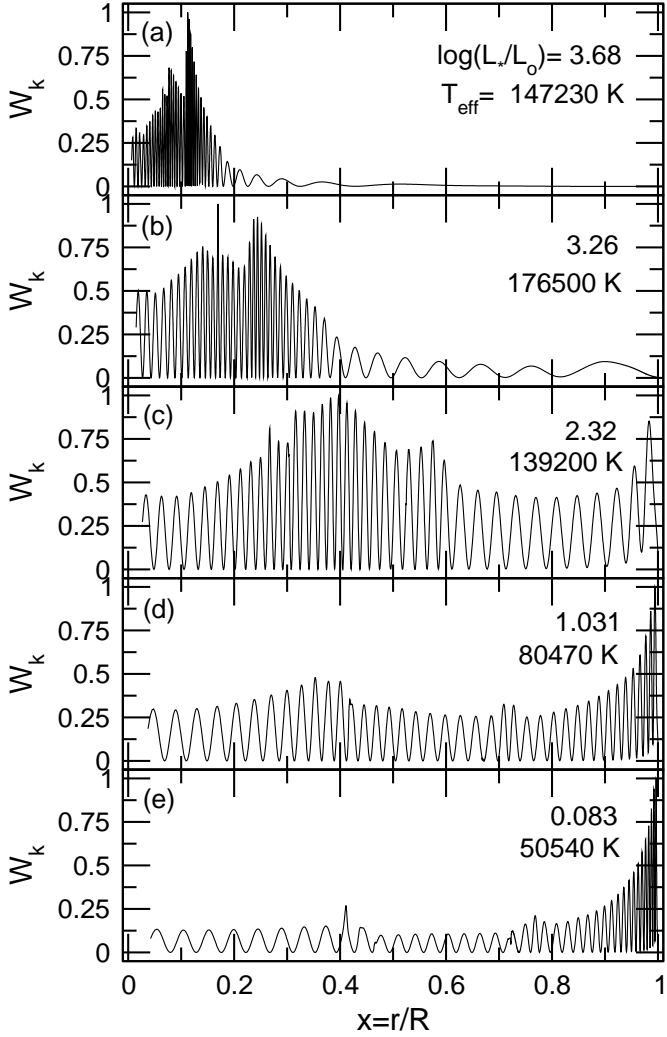}
\caption{The normalized weight function of the mode $g_{40}$ 
vs the  stellar radius  corresponding to $0.5895  M_{\odot}$-models at
different effective  temperatures. The location of  the stellar models
in  the H-R  diagram  is shown  in  Fig. \ref{tracks}.  Note that  the
regions in which $W_k$ adopts appreciable amplitude --- the regions of
period  formation --- move  from the  core toward  the surface  of the
model.}
\label{pesos}
\end{figure}

Note that, because the  very peculiar shape of the Brunt-V\"ais\"al\"a
frequency in  the inner regions of  the star, there  is a considerable
range of intermediate frequencies in which the inner parts of the star
lie  predominantly in the  G-region, and  the outer  parts lie  in the
P-region.   Thus, these intermediate-frequency  modes possess  a mixed
character: they behave  like $g$-modes in the inner  parts of the star
and like  $p$-modes in the  outer parts.  In Fig.   \ref{propa1} these
modes are indicated  with open circles.  Note that  nodes in $y_1$ for
such modes may occur in both the P- and G-regions. In order to get the
radial  order  of  these  modes  we employ  the  scheme  of  Scuflaire
(1974)\footnote{Specifically,  a  mode is  classified  as  $g$ if  the
number of  nodes laying  in the G-region  exceeds the number  of nodes
laying in the P-region ($n_{\rm G} > n_{\rm P}$), and is classified as
$p$  if $n_{\rm  G} <  n_{\rm  P}$. We  then assign  the radial  order
according to $k=  n_{\rm G} - n_{\rm P}$ for  $g$-modes, or $k= n_{\rm
P}  - n_{\rm  G}$  for $p$-modes.}.  In  the figure  modes with  mixed
character  and  some  low   order  ``pure''  $p$-  and  $g$-modes  are
appropriately labeled.

Fig. \ref{propa3} shows a propagation diagram corresponding to a model
characterized by $M_*=  0.5895 M_{\odot}$, $\log(L_*/L_{\odot})= 2.31$
and  $T_{\rm  eff}=  139\,000$  K.   Unlike  the  model  described  in
Fig. \ref{propa1}, this PG 1159  model is already located at the early
WD cooling track --- its effective temperature is marked with an arrow
in panel (b)  of Fig. \ref{per-teff}.  We expect  that the fundamental
properties of  its pulsational spectrum should be  reminiscent to that
of the pulsating WD  stars. In fact, the Brunt-V\"ais\"al\"a frequency
shows larger values  in the outer regions than in the  core, in such a
way that the propagation regions P  and G are markedly separated. As a
result, modes with  mixed character no longer exist,  and the families
of $g$-  and $p$-modes are  clearly differentiated. In this  case, the
radial order  of the modes  is simply the  number of the nodes  of the
radial eigenfunction. Note that for  this model the $g$-modes are free
to propagate  throughout the star,  at variance with the  situation of
the  model described  in  Fig. \ref{propa1},  in  which $g$-modes  are
strongly confined to the core region.

The regions of  the star that most contribute  to the period formation
are inferred from Fig.  \ref{pesos}, which shows the normalized weight
function (Eq.  (\ref{wf}) of the Appendix) of a $g$-mode with $k= 40,
\ell=  1$  in  terms  of the  stellar  radius
corresponding  to  $0.5895  M_{\odot}$-models at  different  effective
temperatures.   The location  of these  models in  the H-R  diagram is
depicted as black dots in Fig. \ref{tracks}.  Note that, for the model
at  $T_{\rm   eff}  \approx  147\,000$  K  (panel   a),  $W_k$  adopts
appreciable amplitudes only in  central regions. Since the behavior of
the weight function  for this mode is representative  of all $g$-modes
with  intermediate and  high radial  order $k$,  we conclude  that the
$g$-mode  pulsation periods  for PG  1159 models  that have  not still
reached the maximum in $T_{\rm eff}$ are mostly determined by the core
regions  of the  star,  irrespective  of the  structure  of the  outer
layers.   Thus, for  models  at high-luminosity  phases the  adiabatic
pulsations properties  reflect the conditions in  the degenerate core.
The  situation for  a model  placed exactly  at the  turning  point in
$T_{\rm  eff}$  is depicted  in  panel  (b)  of Fig.   \ref{per-teff},
corresponding to $T_{\rm eff}=  176\,500$ K and $\log (L_*/L_{\odot})=
3.26$.  Note that, at this stage, the regions of period formation have
extended  towards more  external zones.   By  the time  the model  has
settled  into  its WD  cooling  track,  the  weight function  exhibits
appreciable values throughout the  star, having some amplitude also in
the outer regions,  as displayed in panel (c).   Below $\sim 100\,000$
K, the weight  function adopts its maximum value  in the outer layers,
as can be  seen in panels (d) and (e). At  these stages, the adiabatic
periods  are mainly  determined by  the outer  envelope,  although the
inner regions of  the star also contribute to  establishing the period
of oscillation. Our findings are in full agreement with the results of
the  early computations  of Kawaler  et al.  (1985) ---  for instance,
compare our Fig. \ref{pesos} with their figure 5.

\subsection{Asymptotic period spacing}
\label{asymp}

In the asymptotic limit of  very high radial order $k$ (long periods),
the periods of $g$-modes of a chemically homogeneous stellar model for
a given degree $\ell$ and  consecutive $k$ are separated by a constant
period  spacing  $\Delta \Pi_{\ell}^{\rm  a}$,  given  by (Tassoul  et
al. 1990):

\begin{equation}
\Delta \Pi_{\ell}^{\rm a}=  \frac{\Pi_0}{\sqrt{\ell(\ell+1)}}= 
\frac{2 \pi^2}{\sqrt{\ell(\ell+1)}} \left[ \int_{x_1}^{x_2} \frac{N}{x} 
dx\right]^{-1}.
\label{asymptotic}
\end{equation}

\noindent Note that $\Pi_0$  (and $\Delta \Pi_{\ell}^{\rm a}$) is a 
function  of  the  structural  properties  of  the  star  through  the
Brunt-V\"ais\"al\"a frequency. The integral is taken over the $g$-mode
propagation region. While for  {\it chemically homogeneous} models the
asymptotic  formula  (\ref{asymptotic})  constitutes  a  very  precise
description  of their  pulsational properties,  the  computed $g$-mode
period spacings ($\Delta \Pi_{k} \equiv
\Pi_{k+1}-\Pi_{k}$) in  {\it chemically stratified}  PG 1159
models  show appreciable  departures  from uniformity  caused by  mode
trapping  (see next  Section). However,  the average  of  the computed
period  spacings  $\overline{\Delta \Pi_{k}}$  is  very  close to  the
asymptotic  period spacing $\Delta  \Pi_{\ell}^{\rm a}$  when averaged
over several modes.

\begin{figure}
\centering
\includegraphics[clip,width=250pt]{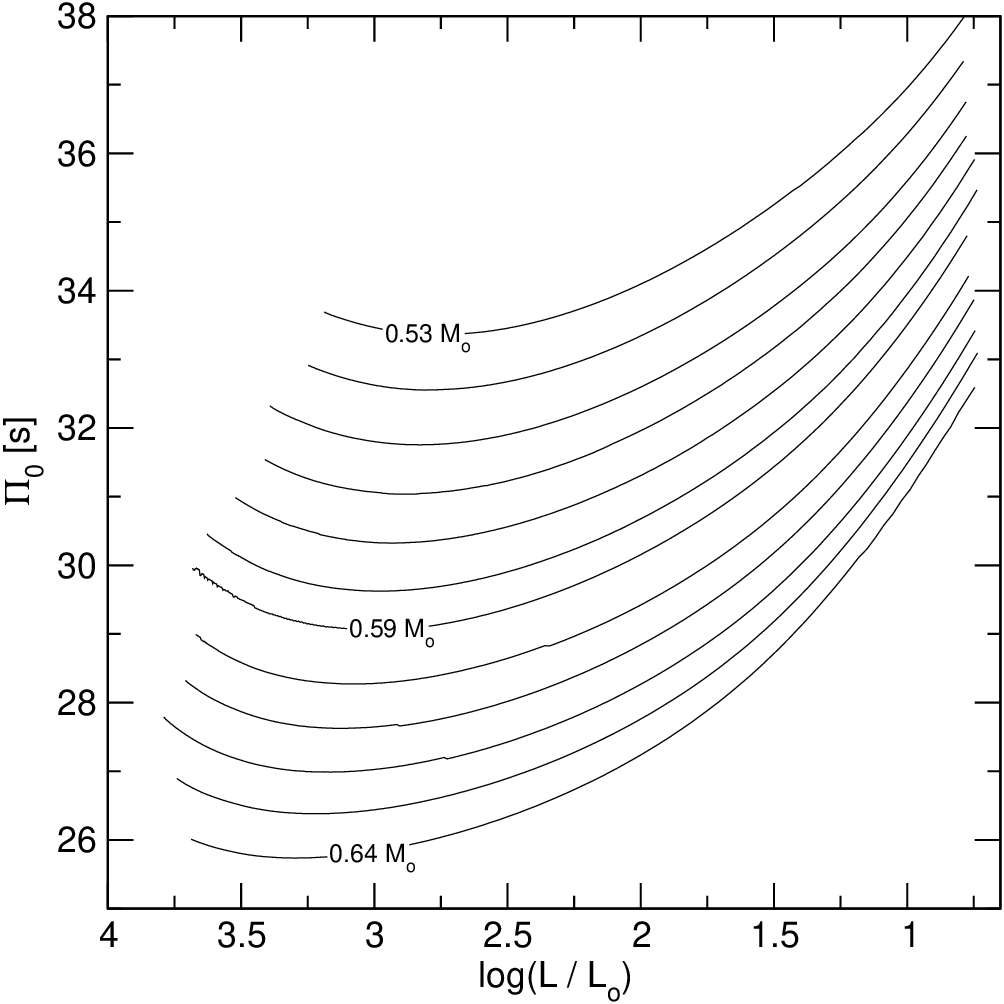}
\caption{The evolution of $\Pi_0$ in terms of stellar luminosity for 
evolutionary  models of  different  stellar masses  ($0.53,\cdots,0.64
M_{\odot}$  from top  to  bottom).  As  in  Figs. \ref{grav-teff}  and
\ref{tracks}, only results corresponding to models whose structure has
completely forgotten the unphysical  transitory stage (after we switch
the  stellar mass  value; see  toward the  end of  Sect.  \ref{2}) are
shown.}
\label{asint}
\end{figure}

The evolution  of $\Pi_0$  vs the stellar  luminosity for  models with
different stellar  masses is displayed in Fig.   \ref{asint}.  In line
with  previous works  (see,  e.g., KB94),  our  results indicate  that
$\Pi_0$  increases with  decreasing stellar  mass. In  particular, the
value  of $\Pi_0$  exhibits an  increment of  about $6-8$  s  when the
stellar mass decreases about $0.1  M_{\odot}$. This is due to the fact
that the  asymptotic period spacing  is inversely proportional  to the
integral  of $N$ (Eq.~\ref{asymptotic}).   As we  go to  lower stellar
masses, the  Brunt-V\"ais\"al\"a frequency decreases  and consequently
$\Pi_0$ goes up.

Fig. \ref{asint} shows that $\Pi_0$  is also a function of the stellar
luminosity, although the dependence is much weaker than the dependence
on   the   stellar  mass.    As   documented   by   the  figure,   for
$\log(L_*/L_{\odot}) \lesssim 3.25-2.75$  --- depending on the stellar
mass   ---  the   period   spacing  increases   with  decreasing   the
luminosity. This fact  can be explained on the basis  that, due to the
increasing   degeneracy  in   the  core   as  the   star   cools,  the
Brunt-V\"ais\"al\"a  frequency  gradually  decreases, causing  a  slow
increment in the magnitude of $\Pi_0$.  Finally, since in this work we
do not consider  different helium layer mass values,  we have not been
able  to  assess the  possible  dependence  of  the asymptotic  period
spacing on $M_{\rm  He}$.  However, this dependence is  expected to be
very weak (see KB94).

\subsection{Mode-trapping properties}
\label{modetrapping}

As  we  have  mentioned  before,  the period  spectrum  of  chemically
homogeneous  stellar  models is  characterized  by  a constant  period
separation, given very accurately by Eq. (\ref {asymptotic}). However,
current evolutionary  calculations predicts that PG 1159  stars --- as
well as (hydrogen-rich) DA and (helium-rich) DB WD stars --- must have
composition  gradients in  their  interiors, something  that also  the
observation indicates.  The presence of  one or more narrow regions in
which the  abundances of nuclear species are  rapidly varying strongly
modifies the  character of the  resonant cavity in which  modes should
propagate as standing waves --- the propagation region.  Specifically,
chemical interfaces act like reflecting boundaries that partially trap
certain  modes, forcing them  to oscillate  with larger  amplitudes in
specific  regions ---  bounded  either  by two  interfaces  or by  one
interface  and the  stellar centre  or  surface ---  and with  smaller
amplitudes outside of  that regions. The requirement for  a mode to be
trapped  is that the  wavelength of  its radial  eigenfunction matches
with  the spatial  separation between  two interfaces  or  between one
interface  and  the  stellar   centre  or  surface.   This  mechanical
resonance,  known as  {\it mode  trapping},  has been  the subject  of
intense study in the context of stratified DA and DB WD pulsations ---
see, e.g., Brassard  et al.  (1992); Bradley et  al. (1993); C\'orsico
et al. (2002).  In the field  of PG 1159 stars, mode trapping has been
extensively explored  by KB94;  we refer the  reader to that  work for
details.

There are  (at least) two  ways to identify  trapped modes in  a given
stellar model.  First, we can consider the oscillation kinetic energy,
$K_{k}$.  We  stress that, because the amplitude  of eigenfunctions is
arbitrarily normalized  at the model surface in  our calculations (see
the Appendix), the  values of the kinetic energy are  useful only in a
{\it relative} sense.  $K_{k}$ is  proportional to the integral of the
squared  amplitude of  eigenfunctions,  weighted by  the density  (see
Eq. \ref{kinetic}  in the Appendix).   Thus, modes propagating  in the
deep interior of the star, where densities are very high, will exhibit
larger  values than modes  which are  oscillating in  the low-density,
external regions.   When only a single chemical  interface is present,
modes can  be classified as modes  trapped in the  outer layers, modes
confined in  the core  regions, or simply  ``normal modes''  --- which
oscillate everywhere in  the star --- characterized by  low, high, and
intermediate  $K_{k}$  values,   respectively  (see  Brassard  et  al.
1992). This  rather simple picture  becomes markedly complex  when the
stellar model  is characterized by {\it  several} chemical composition
gradients ---  see C\'orsico et  al.  (2002) for  the case of  ZZ Ceti
models.

\begin{figure}
\centering
\includegraphics[clip,width=250pt]{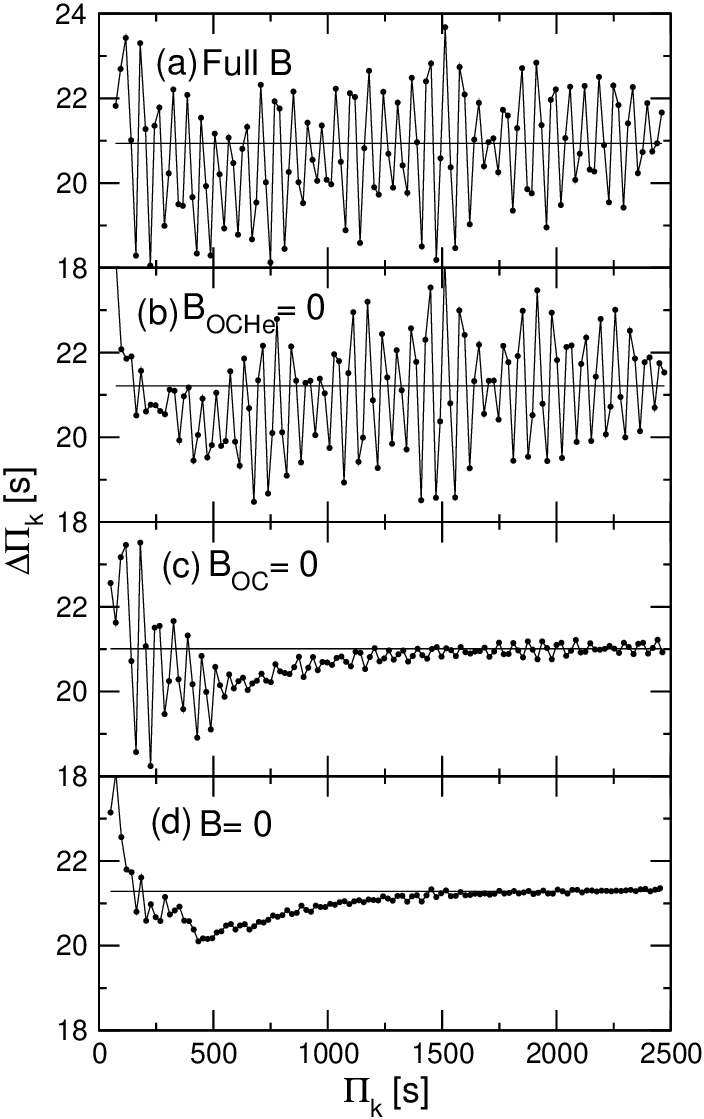}
\caption{The forward period spacing $\Delta \Pi_{k}$ vs period 
$\Pi_{k}$ for dipole ($\ell=  1$) modes. The asymptotic period spacing
$\Delta  \Pi_{\ell}^{\rm a}$  as  given by  Eq. (\ref{asymptotic})  is
depicted with  a thin  horizontal line. Panel  (a) corresponds  to the
same PG 1159 model considered in Fig.  \ref{profile}. Remainder panels
correspond to  the same stellar model  but for the  situation in which
the Ledoux  term has been artificially suppressed  in specific regions
of the  model: (b)  $B= 0$ at  $M_r/M_* \approx  0.96$, (c) $B=  0$ at
$M_r/M_* \approx 0.4-0.6$, and (d) $B=0$ everywhere in the star.}
\label{deltap1}
\end{figure}

A second  --- and more important  from an observational  point of view
--- consequence of mode trapping is that the period spacing $\Delta
\Pi_{k}$,  when   plotted  in  terms  of   the  pulsation  period
$\Pi_{k}$,  exhibits strong  departures  from uniformity.   It is  the
period difference between an observed mode and adjacent modes ($\Delta
k= \pm 1$)  that is the observational diagnostic  of mode trapping ---
at variance  with $K_{k}$, whose  value is very difficult  to estimate
only from  observation.  For stellar models characterized  by a single
chemical   interface,  local  minima   in  $\Delta   \Pi_{k}$  usually
correspond  directly to  modes trapped  in the  outer  layers, whereas
local maxima  in $\Delta \Pi_{k}$  are associated to modes  trapped in
the core region.

We now turn out to the specific case of our PG 1159 pre-WD models.  As
Fig.  \ref{profile} shows, they  exhibit several  chemical interfaces,
some associated  with the various  steps in the $^{16}$O  and $^{12}$C
profile at  the core  region, and a  single, more  external transition
region  in which oxygen,  carbon and  helium are  continuously varying
(see  Fig.  \ref{profile}).  As  stated  in  Sect.  \ref{code},  these
composition    gradients    produce    pronounced   peaks    in    the
Brunt-V\"ais\"al\"a  frequency ---  through the  Ledoux term  --- that
strongly disturb the structure of the period spectrum.
                                        
The influence of the chemical composition gradients inside our PG 1159
models on their period spectra is clearly shown in panel (a) of Fig.
\ref{deltap1}, in which the $\ell= 1$ period 
spacing  is plotted  in terms  of the  periods for  a  reference model
characterized by  a stellar mass  of $0.5895 M_{\odot}$,  an effective
temperature of $139\,000$ K and a luminosity of $204.2 L_{\odot}$. The
plot shows very rapid variations of $\Delta \Pi_{k}$ everywhere in the
period spectrum, with  ``trapping amplitudes'' up to about  6 s and an
asymptotic period  spacing of $\approx  20.94$ s.  The  rather complex
period-spacing  diagram shown  by  Fig.  \ref{deltap1}  is typical  of
models  characterized by  several  chemical interfaces.   In order  to
isolate  the effect  of  each chemical  composition  gradient on  mode
trapping,  we follow  the procedure  of  Charpinet et  al. (2000)  for
models of sdB  stars --- see also Brassard et al.  (1992) for the case
of  ZZ  Ceti  stars.    Specifically,  we  minimize  ---  although  no
completely  eliminate --- the  effects of  a given  chemical interface
simply  by forcing  the Ledoux  term $B$  to be  zero in  the specific
region of  the star in which  such interface is located.  In this way,
the resulting mode trapping will be only due to the remainder chemical
interfaces.   In  the  interest  of  clarity, we  label  as  ``$B_{\rm
OCHe}$'' the contribution to $B$  due to the O/C/He chemical interface
at $M_r/M_*
\approx 0.96$, and ``$B_{\rm OC}$'' the contribution to $B$ associated
with the O/C chemical interface at $M_r/M_*
\approx 0.4-0.6$  --- see inset of Fig.  \ref{profile}. Specifically, we
have  recomputed  the  entire   $g$-mode  period  spectrum  under  the
following assumptions: (1) $B_{\rm OCHe}=  0$ and $B_{\rm OC} \neq 0$,
(2) $B_{\rm OCHe} \neq 0$ and  $B_{\rm OC}= 0$, and (3) $B_{\rm OCHe}=
0$ and $B_{\rm OC}= 0$ ($B=0$  in all regions).  The results are shown
in panels (b),  (c) and (d) of Fig.   \ref{deltap1}, respectively.  By
comparing  the different  cases illustrated,  an  important conclusion
emerges from this figure: the chemical transition region at $M_r
\approx 0.96  M_*$ is responsible for the  non-uniformities in $\Delta
\Pi_{k}$ only for $\Pi_{k}
\lesssim 500 $ s (panel c), whereas the chemical composition gradients
at the core region ($M_r \approx 0.4-0.6 M_*$) cause the mode-trapping
structure in the  rest of the period spectrum (panel  b).  When $B= 0$
everywhere inside  the model (panel d), the  period-spacing diagram is
characterized  by the  absence of  strong features  of  mode trapping,
although  some   structure  remains,  in  particular   for  low  order
modes. Note  the nice agreement between the  numerical computations of
$\Delta \Pi_{k}$ and the predictions of the asymptotic theory given by
Eq. (\ref{asymptotic}) for the limit $k \gg 1$.

From the  above discussion,  it is clear  that {\it  the mode-trapping
features  characterizing  our PG  1159  models  are  inflicted by  the
stepped shape  of the carbon/oxygen  chemical profile at the  core ---
left by  prior convective  overshooting--- at least  for the  range of
periods observed in  GW Vir stars}.  On their  hand, the more external
chemical transition  has a minor  influence, except for the  regime of
short periods. We mention that  this situation is more evident for the
more massive PG 1159 models than for the less massive ones.

\begin{figure}
\centering
\includegraphics[clip,width=250pt]{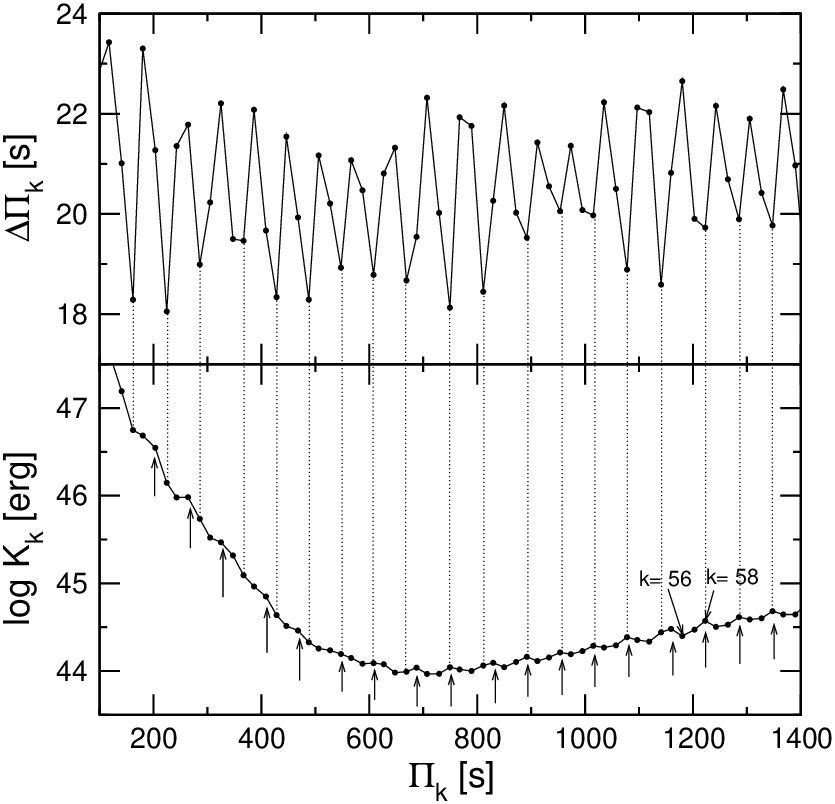}
\caption{Period-spacing diagram and kinetic energy 
distribution  (upper and lower  panel, respectively)  for the  same PG
1159 model  considered in panel (a) of  Fig.  \ref{deltap1}.  Vertical
dotted  lines   link  local  minima  in  $\Delta   \Pi_{k}$  with  the
corresponding values of $K_{k}$.  Maximum values of kinetic energy are
indicated with arrows. Also labeled are the modes with $k= 56$ and $k=
58$, to be analyzed in Fig. \ref{afu-weight}.}
\label{dp-ekin}
\end{figure}

The findings outlined  above are clearly at odd  with previous results
reported  by  KB94.   Indeed,   these  authors  have  found  that  the
mode-trapping properties of  their PG 1159 models are  fixed mainly by
the outer O/C/He transition region, to  a such a degree that they have
been able to employ mode-trapping signatures as a sensitive locator of
this transition region.   The origin of this discrepancy  can be found
in  the  details  of  the  input  physics employed  in  build  up  the
background stellar models. Of particular interest here is the presence
of a  much less pronounced chemical  transition in the  outer parts of
the C/O  core of the KB94  models, as compared with  the rather abrupt
chemical gradients at $M_r  \approx 0.4-0.6 M_*$ characterizing our PG
1159  models.   If  we   artificially  minimize  the  effect  of  this
transition region  --- by setting  $B_{\rm OC}= 0$ ---  we immediately
recover the  results of KB94.   Indeed, the period-spacing  diagram of
panel (c) in  our Fig. \ref{deltap1} looks very  similar to that shown
in Fig.  3 of  KB94, corresponding to  $M_{\rm He}/M_*=  0.00622$.  We
note,  however, that the  ``trapping cycle''  --- the  period interval
between two period-spacing minima ---  of our modified model (c) is of
$\approx 70$  s, whereas the value  of the KB94 model  is $\approx 95$
s. This difference is due mainly to the fact that our model has a more
massive helium envelope ($M_{\rm He}/M_*= 0.00881$) than that of KB94.
 
By means of a simple  numerical experiment we have identified the main
source of  mode trapping as  due to the step-like  chemical transition
region  located  at  the  core.    We  now  return  to  panel  (a)  of
Fig. \ref{deltap1}, and note that there is a kind of ``beating'' which
modulates  the trapping amplitude.   We also  note that  this striking
feature (seen in all of our models) persists even in the case in which
the effect of  the O/C/He transition is artificially  minimized, as it
is illustrated in panel (b).  We have found that the beating exhibited
by   the  period-spacing   distribution   is  due   to  the   combined
mode-trapping effects caused by the  various steps in the O/C chemical
profile in  the core.  In  fact, by performing period  computations in
which only  the largest peak of  the Ledoux term at  $M_r \approx 0.58
M_*$  (see inset of  Fig.  \ref{profile})  is considered,  the beating
effect virtually  vanishes, and the trapping  amplitude becomes nearly
constant.

Finally, to understand in  more detail the mode-trapping properties of
our  models,   we  resort  to   the  kinetic  energy  of   modes.   In
Fig. \ref{dp-ekin} we  plot the period spacing and  the kinetic energy
distribution  (upper and lower  panel, respectively)  for the  same PG
1159 model considered in panel  (a) of Fig. \ref{deltap1}.  We connect
with  dotted  lines  each  minimum  in  $\Delta  \Pi_{k}$  with  their
corresponding value  of $K_{k}$. We note  that in spite  of the strong
non-uniformities exhibited by $\Delta  \Pi_{k}$, the $K_{k}$ values of
adjacent modes seem to be  not very different amongst them. However, a
closer inspection  of the figure reveals  that in most of  the cases a
minimum in $\Delta \Pi_{k}$ corresponds to a local maximum in $K_{k}$.
Our results strongly suggest that {\it modes associated with minima in
the period-spacing diagram usually correspond to maxima in the kinetic
energy distribution. They should  be modes characterized by relatively
high  amplitude of their  eigenfunctions and  weight functions  in the
high-density environment characterizing the stellar core}.

\begin{figure}
\centering
\includegraphics[clip,width=250pt]{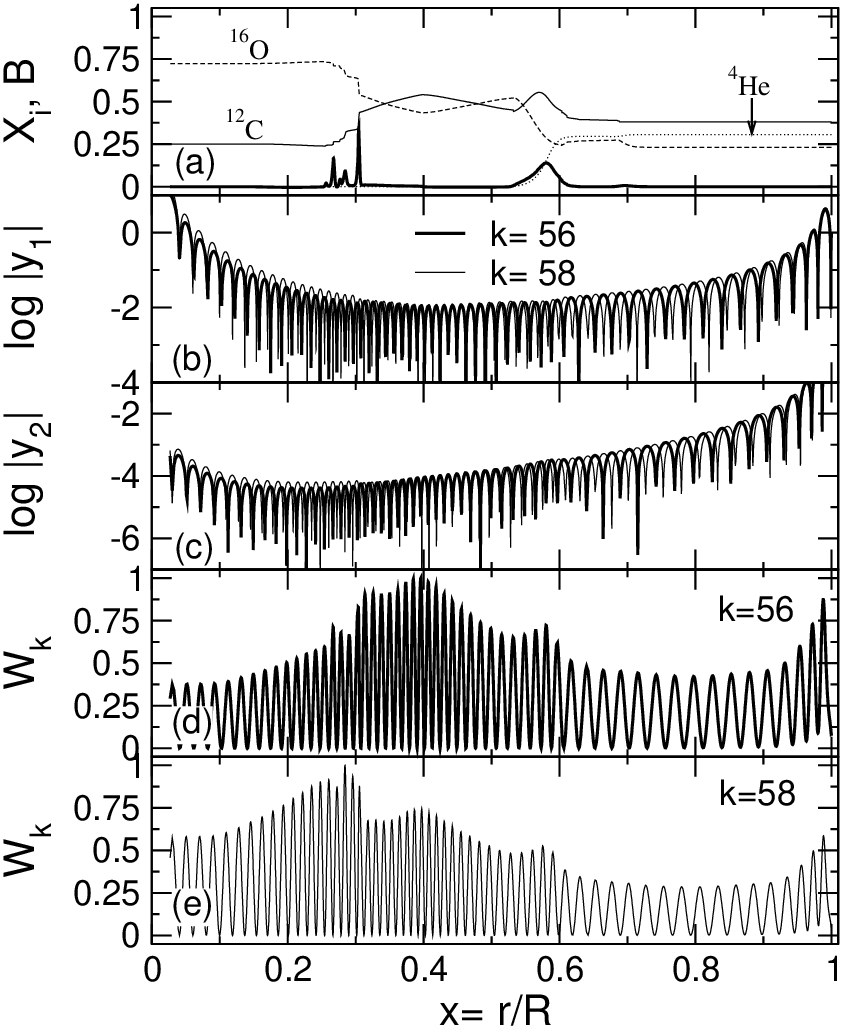}
\caption{The spatial run of the modulus of the eigenfunctions $y_1$ 
and $y_2$, shown  in panels (b) and (c),  respectively, and the weight
function $W_k$  depicted in panels  (d) and (e), for  dipole pulsation
modes with $k= 56$ (thick  line) and $k= 58$ (thin line) corresponding
to the  same PG 1159 model  analyzed in Fig.   \ref{dp-ekin}.  We also
include in panel  (a) the Ledoux term $B$ and  the abundance (by mass)
of $^{16}$O, $^{12}$C and $^{4}$He.}
\label{afu-weight}
\end{figure}

This is precisely what we have  found after a careful examining of the
eigenfunctions at the deepest regions  of the core.  In panels (b) and
(c) of  Fig. \ref{afu-weight}  we show the  logarithm of  the absolute
value of the eigenfunctions $y_1$ and $y_2$ (see Eq. \ref{y1y2} of the
Appendix  for  their  definition),   respectively,  in  terms  of  the
dimensionless  radius $x=r/R$,  for a  mode having  a local  maxima in
kinetic energy  ($k= 58$,  thin line)  and for a  mode having  a local
minima in $K_{k}$ ($k= 56$,  thick line). Note that below $r/R \approx
0.3$, the  eigenfunctions of the  $k= 58$ mode have  larger amplitudes
than that of the  $k= 56$ mode, explaining why the $k=  58$ mode has a
relatively large  oscillation kinetic energy.  Modes like  the $k= 58$
one are partially  trapped in the core region,  below the O/C chemical
interface  (see panel a).   As stated  before, these  modes correspond
generally to minima in the  period-spacing diagram.  In panels (d) and
(e) of  Fig.  \ref{afu-weight} the normalized  weight function ($W_k$)
is displayed for  the $k= 56$ and $k= 58$  modes, respectively.  As we
mentioned  previously,  the  relative  values  of $W_k$  for  a  given
pulsation mode  provide information about the  specific regions inside
the  star  that  most   contribute  to  the  period  formation.   From
Fig. \ref{afu-weight} we note that for the $k= 58$ mode the maximum of
$W_k$ is  identified with the  O/C chemical interface at  $r/R \approx
0.3$,  and the  largest amplitude  portion of  the weight  function is
located in the deepest regions  of the core ($r/R \lesssim 0.3$).  For
values  of  $r/R$ slightly  larger  than  $0.3$,  the weight  function
abruptly diminishes and then reaches a secondary maximum in the region
in  which $^{12}$C becomes  more abundant  than $^{16}$O.   Above this
region $W_k$ exhibit other  secondary maximum at the O/C/He transition
zone  ($r/R \approx  0.58$), and  above  of this  the weight  function
adopts rather  lower values ($W_k  \lesssim 0.3$). Thus, the  shape of
the weight  function for the $k=  58$ mode strongly  suggests that the
deepest  regions of  the core  have the  larger impact  on  the period
formation  of this  mode.  This  conclusion is,  not  surprisingly, in
complete agreement with the fact  that this mode is characterized by a
relatively  high  value  of   the  oscillation  kinetic  energy.   The
situation is markedly  different for the case of the  $k= 56$ mode, as
it is clearly demonstrated by panel  (d) of the figure. For this mode,
most regions  of the star  appreciably contribute to the  formation of
its period.  In particular,  the most important contributions occur in
regions located at $0.2 \lesssim r/R \lesssim 0.6$.

\begin{figure}
\centering
\includegraphics[clip,width=250pt]{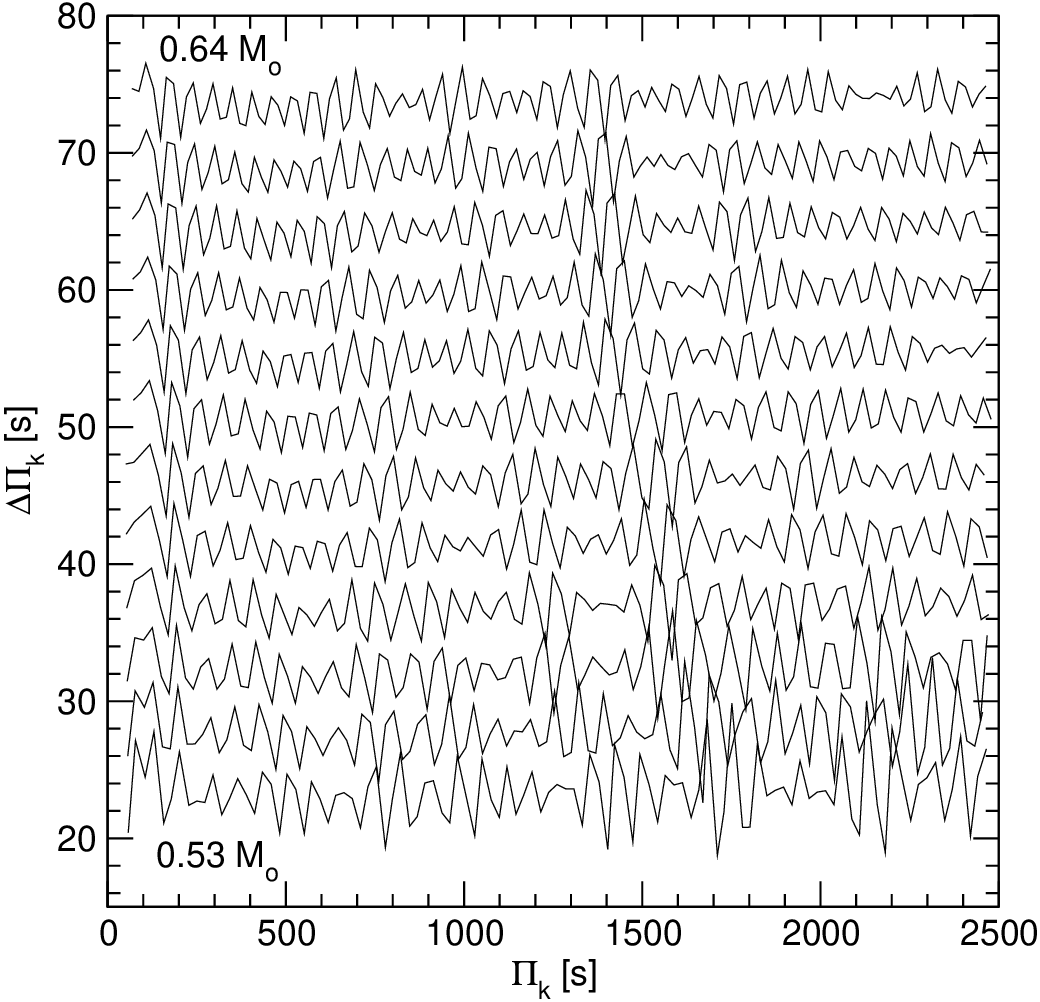}
\caption{The period-spacing  distributions for 
PG 1159 models with stellar masses, from bottom to top: $0.53, 0.54,
\ldots, 0.64 M_{\odot}$ at $T_{\rm eff}
\approx 140\,000$ K. In the interests of clarity, each curve has
been  arbitrarily  shifted  upwards,  starting from  the  lowest  one,
corresponding to  the model with  $M_*= 0.53 M_{\odot}$.   Models with
stellar  masses other than  $0.5895$-\msun correspond  to evolutionary
stages after  the unphysical transitory phase (induced  by the scaling
of the stellar mass) has long vanished.}
\label{dp-masas}
\end{figure}

\subsection{The effect of changes in the stellar mass}
\label{mass}

In the  above discussion  we have been  able to identify  the physical
origins and nature of mode trapping in our PG 1159 models. We wish now
to  explore the  effect  of changing  the  stellar mass  $M_*$ on  the
period-spacing diagrams and mode trapping. As mentioned previously, we
have employed in  this project model sequences with  stellar masses in
the ranges  $0.53-0.58$ \msun~ and  $0.60-0.64$ \msun~ with a  step of
0.01  \msun~, in  addition  to the  0.5895-\msun~ sequence  previously
described.   We recall  that those  sequences were  obtained  from the
0.5895-\msun~  sequence  by  artificially  changing the  stellar  mass
according the procedure outlined in  \S \ref{2}. Thus, we caution that
the  results derived  in this  section  could change  if the  complete
evolution  of  progenitor stars  with  different  stellar masses  were
accounted for. From the asymptotic theory outlined in \S
\ref{asymp}, we  know that $\Delta \Pi_{\ell}^{\rm  a}$ decreases when
the stellar mass increases, as it is evident from Fig.
\ref{asint}. As mentioned previously, the asymptotic period spacing is
very   close  to  the   average  of   the  computed   period  spacings
$\overline{\Delta \Pi_{k}}$.  As a result, the average of the computed
period spacing also must decreases  as the stellar mass increases.  In
fact, $\overline{\Delta \Pi_{k}}$ decreases  from $23.19$ s to $18.69$
s when  we increase  the stellar mass  from $0.53 M_{\odot}$  to $0.64
M_{\odot}$ for a fixed $T_{\rm eff} \approx 140\,000$ K.

We also expect variations in  the amplitude and cycle of trapping when
we consider changes  in the stellar mass.  In  Fig.  \ref{dp-masas} we
depict period-spacing diagrams for PG 1159 models with values of $M_*$
covering the complete  set of stellar masses considered  in this work,
and for a  fixed effective temperature of $\approx  140\,000$ K.  Note
that  the   trapping  amplitude   decreases  when  the   stellar  mass
increases. In fact, the maximum  of the trapping amplitude is of about
$10$ s for $0.53 M_{\odot}$, whereas it assumes a value of $\approx 5$
s, at most,  for $0.64 M_{\odot}$. This is a  direct consequence of an
increased electron degeneracy in the  core of our more massive models,
which  in  turn  produces  a  weakening  in  the  efficiency  of  mode
trapping. Other striking feature observed in Fig.
\ref{dp-masas} is
the decrease in the period of the trapping cycle and the shift of mode
trapping features  to lower periods as  we go to larger  values of the
stellar mass. Specifically, the portions of the period-spacing diagram
centered at $\approx 500, \approx 900, \approx 1400, \approx 1700$ and
$\approx 2200$  s for  the model with  $M_*= 0.53 M_{\odot}$,  move to
regions centered  at $\approx 100, \approx 700,  \approx 1000, \approx
1400$ and  $\approx 1900$  s, respectively, for  the model  with $M_*=
0.64 M_{\odot}$.   This effect  can be understood  on the  basis that,
while the O/C  chemical structure remains at the  same $M_r/M_*$ value
inside the model when we  consider higher stellar masses, its location
in terms of  the radial coordinate ($r$) shifts  away from the stellar
centre, from $r/R \approx 0.22$ for $M_*= 0.53 M_{\odot}$ to $r/R
\approx 0.33$  for $M_*=  0.64 M_{\odot}$. A  similar effect  has been
found by  C\'orsico et al. (2005)  in the context  of crystallizing ZZ
Ceti models, by considering a fixed  value of $M_*$ but with the inner
turning point  of the $g$-mode eigenfunctions  --- the crystallization
front --- moving outwards.

\subsection{The effect of changes in the effective temperature}
\label{teff}

In  closing  this section,  we  shall  explore  the evolution  of  the
mode-trapping properties as our PG 1159 evolve to cooler temperatures.
As  we  mentioned  earlier,  the  asymptotic  period  spacing  $\Delta
\Pi_{\ell}^{\rm a}$  and the average  of the computed  period spacings
$\overline{\Delta  \Pi_{k}}$  increase  when  the  stellar  luminosity
decreases (see Fig.  \ref{asint}).  When the star evolves along the WD
cooling   track,  a  decreasing   effective  temperature   is  usually
associated to a  decreasing stellar luminosity.  As a  result, we also
expect   that  $\Delta   \Pi_{\ell}^{\rm  a}$   and  $\overline{\Delta
\Pi_{k}}$ increase  as cooling proceeds.   This trend is  confirmed by
our  pulsation  computations.   In fact,  $\overline{\Delta  \Pi_{k}}$
increases  from $20.22$  s to  $\approx  24.12$ s  when the  effective
temperature is decreased  from $170\,000$ K to $70\,000$  K in a model
with  $0.5895   M_{\odot}$.   The   explanation  to  this   effect  is
straightforward. As the star  cools their core becomes more degenerate
and the Brunt-V\"ais\"al\"a frequency decreases because $\chi_{\rm T}$
goes  down  (see Eq.   \ref{bruntva}).   As  a  result, the  pulsation
periods become longer  and thus the period spacing  and the average of
the period spacing must increase.

\begin{figure}
\centering
\includegraphics[clip,width=250pt]{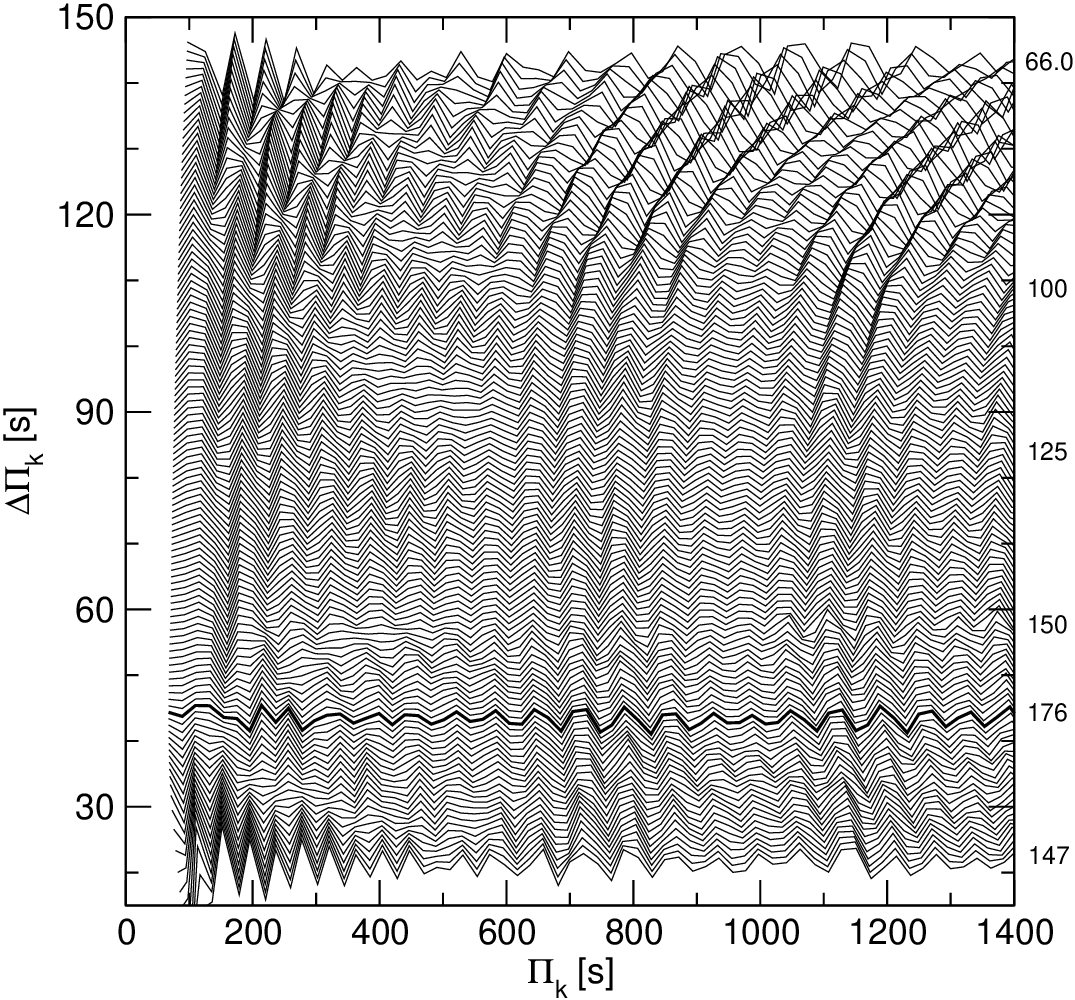}
\caption{Period-spacing  distributions for 
0.5895-\msun~ PG 1159 models  at different effective temperatures.  In
the interests of clarity, each  curve has been shifted upwards, except
for the  lowest one.  In a few  cases we have included  numbers at the
right side  of the plot, corresponding  to the $T_{\rm  eff}$ value of
the  models  (in  units  of   kK).  In  particular,  the  thick  curve
corresponds to  the highest value of effective  temperature reached by
the star, $T_{\rm eff} \sim 176$ kK.}
\label{dp-teff}
\end{figure}

Now, we want to see the effect of changes in the effective temperature
on the trapping cycle and amplitude.  To have a global picture of what
happens  when   the  effective  temperature  changes,   we  depict  in
Fig.   \ref{dp-teff}  the   evolution  of   the  period   spacing  for
0.5895-\msun~  models,  shown   at  different  effective  temperatures
(indicated at  the right side  of the plot)  starting from a  stage in
which the model is still approaching its maximum $T_{\rm eff}$ (lowest
curve)  until to a  phase in  which the  object is  already on  the WD
cooling track (upper  curve). Note that the scale at  left of the plot
has sense only for the  lowest curve; the remaining curves are shifted
upward  with purposes of  clarity. The  figure emphasizes  the notable
changes that  undergoes the period-spacing structure  as the effective
temperature  varies.   Note that  there  exist  two clearly  different
trends. In stages in which the effective temperature increases (before
the  star  reaches its  hot  turning  point  at $T_{\rm  eff}  \approx
176,000$  K)  the  mode-trapping  features slowly  move  toward  short
periods. This  is due to the  fact that pulsation  periods decrease in
response of the overall warming of the star, and that the outer layers
contract. When the star has  already passed through its hottest phase,
the  opposite behavior  is  exhibited by  the mode-trapping  features.
Indeed, maxima  and minima values of  the period-spacing distributions
move to  higher periods when the effective  temperature diminishes, in
particular below $T_{\rm eff}
\approx 100\,000$ K. Finally, we note that
the  trapping  amplitude  increases  when  the  effective  temperature
decreases.

\section{Application to real stars: asteroseismology of 
variable GW Vir stars}
\label{4}

In  this  section  we  shall  attempt to  infer  the  main  structural
parameters (that is, $M_*$ and  $T_{\rm eff}$) of several GW Vir stars
by  employing  the  results   of  our  extensive  period  computations
described in previous  sections. Due to the fact  that in this project
we have not been able to compute realistic PG 1159 models with stellar
masses other than $0.5895 M_{\odot}$ in the low-gravity phases (before
the star reaches  the ``knee'' at the highest  value of $T_{\rm eff}$;
see Fig.  \ref{grav-teff}), we  have limited ourselves to make seismic
inferences only for the high-gravity members of the GW Vir class, that
is, the naked  --- stars without surrounding nebulae  --- variables PG
1159-035, PG 2131+066, PG 1707+427,  and PG 0122+200.  We mention that
the results  described below are  conditional on the  assumption under
the  sequences of  different  stellar masses  have  been created  (see
Sect. \ref{2}). As  it is well known, changes in  the thickness of the
surface helium-rich  layer and in  the surface composition  affect the
trapping  cycle and  amplitude  (see KB94).  Thus,  by adopting  these
quantities as free  adjustable parameters it is possible  to perform a
fine tunning of the trapping  cycle and the trapping amplitude to best
match the period  spectrum of a given star. In spite  of this, in this
work we prefer to keep fixed the thickness of the helium layer and the
surface  composition  by  adopting  for these  parameters  the  values
predicted by  our evolutionary computations.   It is worth  to mention
that  all of these  naked GW  Vir stars  exhibit very  similar surface
abundances $(X_{\rm H}, X_{\rm He}, X_{\rm C}, X_{\rm O})
\approx   (0.00,  0.33,  0.50,   0.17)$.  Our   PG  1159   models  are
characterized by  surface abundances of (0.000,  0.306, 0.376, 0.228),
and thus, they are very  appropriate for our purposes. Before going to
seismic applications  of our bank  of periods, we summarize  the basic
properties of the cited GW Vir stars below.

The four naked  GW Vir stars have higher gravity  ($\log g \gtrsim 7$)
than the PNNVs, and they are thought to be slightly more evolved. From
a pulsation point  of view, the only difference  between both types of
variable  stars is  that PNNV  stars  in general  pulsate with  longer
periods --- in the  range of about $1000 - 3000$ s  --- than the naked
GW Vir  --- which  pulsate with periods  below 1000  s.  Specifically,
there is a well-defined correlation between the luminosity of the star
and the periods  exhibited: the more luminous the  star is, the longer
the pulsation periods are.

\vskip 2mm
\noindent {\sl -- PG 1159-035:} this is the prototype of the PG 1159 
spectral class, and also the  prototype of the GW Vir pulsators. After
the discovery of its variability  by McGraw et al. (1979), PG 1159-035
became  the target  of  an intense  observational  scrutiny. The  most
fruitful analysis of its light curve  was carried out by Winget et al.
(1991) employing the high-quality data  from the Whole Earth Telescope
(WET; Nather  et al.   1990).  This analysis  showed that  PG 1159-035
pulsate in  more than 100  independent modes with periods  between 300
and  800 s.   By using  20 unambiguously  identified $\ell=  1,  m= 0$
periods between 430 and 817 s,  KB94 determined a mass for PG 1159-035
of  $0.59  M_{\odot}$,  an  effective  temperature  of  $T_{\rm  eff}=
136\,000$ K  and a surface gravity  of $\log g= 7.4$.   In addition, a
stellar luminosity of $195 L_{\odot}$  and a distance of $440$ pc from
the Earth were inferred. On  the other hand, spectroscopic analysis by
Dreizler \& Heber (1998) yield  $T_{\rm eff}= 140\,000$ K and $\log g=
7$, in good agreement with the asteroseismic fit of KB94, but a higher
luminosity of $501 L_{\odot}$.

\vskip 2mm
\noindent {\sl -- PG 2131+066:} Was discovered as a variable star 
by Bond et al. (1984) with periods  of about 414 and 386 s, along with
some other periodicities. On the  basis of an augmented set of periods
from  WET data,  Kawaler et  al.   (1995) considered  a $T_{\rm  eff}=
80\,000$ K  and obtained  a precise mass  determination of  $M_*= 0.61
M_{\odot}$, a  luminosity of  $10 L_{\odot}$ and  a distance  from the
Earth  of $470$ pc.   Spectroscopic constraints  of Dreizler  \& Heber
(1998) give  $M_*= 0.55  M_{\odot}$, $T_{\rm  eff}= 95\,000$  K, $39.8
L_{\odot}$ and  $\log g= 7.5$ for  PG 2131+066.  By  using the updated
Dreizler \& Heber (1998)'s determination of the effective temperature,
Reed et  al. (2000) refined the  procedure of Kawaler  et al.  (1995).
They found $M_*= 0.608 M_{\odot}$,  a luminosity of $26 L_{\odot}$ and
a seismic distance to PG 2131+066 of $668$ pc.

\vskip 2mm
\noindent {\sl -- PG 1707+427:} This star was discovered to be a pulsator 
by  Bond et  al. (1984).   Dreizler \&  Heber (1998)  obtained $T_{\rm
eff}= 85\,000$ K and $\log g=  7.5$, and a stellar mass and luminosity
of $0.54  M_{\odot}$ and  $25 L_{\odot}$, respectively,  were inferred
from their  spectroscopic study. Recently, Kawaler et  al. (2004) have
reported  the  analysis  of  multisite  observations  of  PG  1707+427
obtained  with WET.  Preliminary  model fits  by  using 7  independent
$\ell=1$  modes   with  periods  between  334  and   910  ssuggest  an
asteroseismic  mass  and  luminosity   of  $0.57  M_{\odot}$  and  $23
L_{\odot}$, respectively.

\vskip 2mm
\noindent {\sl -- PG 0122+200:} It is the coolest GW Vir variable, with 
$T_{\rm eff}=  80\,000$ K and  $L_*= 20 L_{\odot}$ (Dreizler  \& Heber
1998).  Spectroscopy indicates a  stellar mass of $0.53 M_{\odot}$ and
a  gravity  of $\log  g=  7.5$.  By  employing  an  analysis based  on
multisite  observations with  WET,  O'Brien et  al.   (1998) report  a
seismic stellar  mass of $0.69 M_{\odot}$, strikingly  higher than the
spectroscopic estimation.   The cooling of  PG 0122+200 appears  to be
dominated by  neutrino looses; this  renders PG 0122+200 as  the prime
target for learning neutrino physics (O'Brien et al. 1998).

\begin{figure}
\centering
\includegraphics[clip,width=250pt]{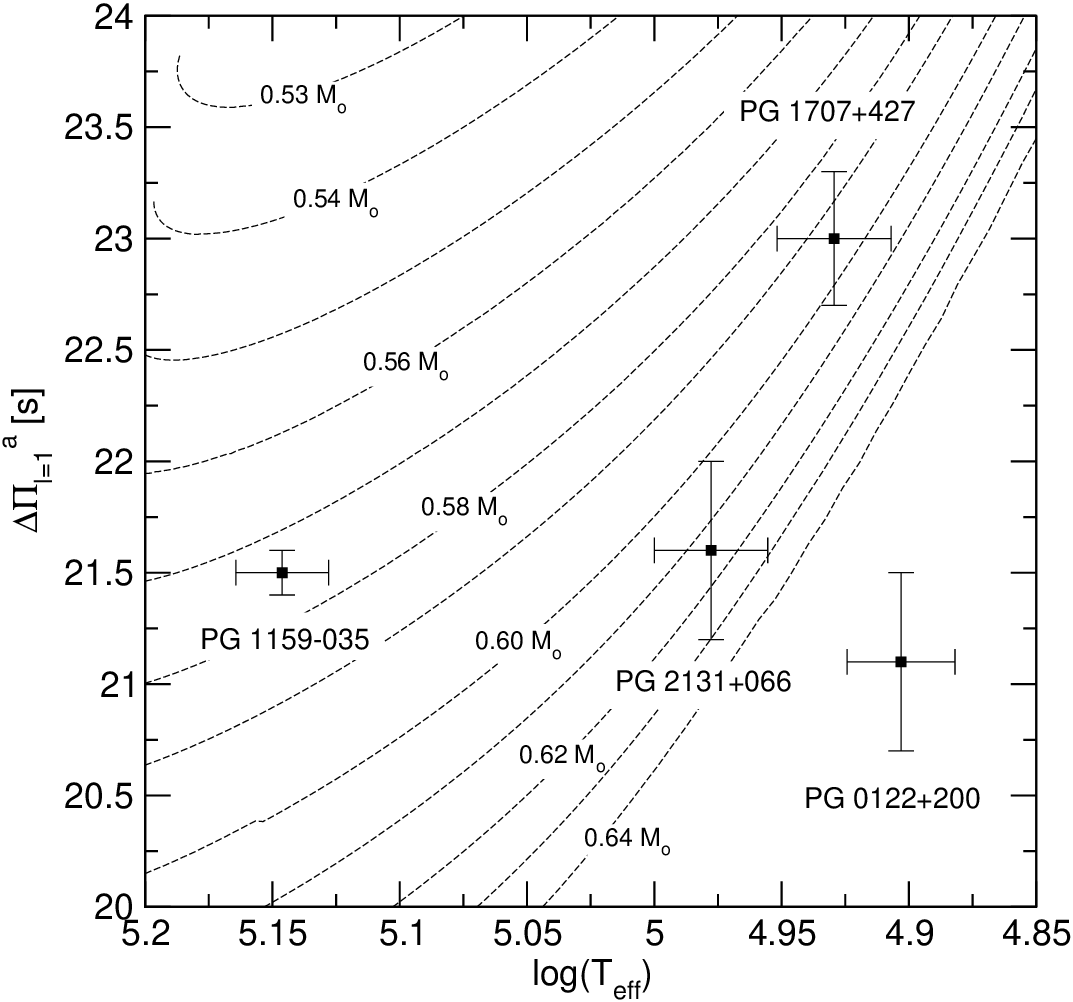}
\caption{The asymptotic period spacing for PG 1159 models with 
stellar  masses  of  $0.53,\cdots,0.64  M_{\odot}$  in  terms  of  the
effective temperature. This figure  is based on Fig. \ref{asint}.  The
observed   mean  period   spacing  $\overline{\Delta   \Pi}$   of  the
high-gravity GW Vir  stars PG 1159-035, PG 2131+066,  PG 1707+427, and
PG 0122+200 ($21.5 \pm 0.1$ s, $21.6  \pm 0.4$ s, $23.0 \pm 0.3$ s and
$21.1 \pm 0.4$  s, respectively) is also shown  (taken from Kawaler et
al. 2004).}
\label{asint2}
\end{figure}

\begin{table*}
\centering
\caption{The values of the stellar mass (in $M_{\odot}$) from spectroscopic 
and asteroseismic studies, and the values obtained in this work from three 
different methods.}
\begin{tabular}{lcccc}
\hline
\hline
Source &  PG 1159$-$035 &  PG 2131+066  & PG 1707+427 &  PG 0122+200 \\
\hline
Spectroscopy     &  $0.54 \pm 0.1^{\rm (a)}$  & $0.55 \pm 0.1^{\rm (a)}$   & $0.54 \pm 0.1^{\rm (a)}$ &  $0.53 \pm 0.1^{\rm (a)}$ \\ 
Asteroseismology  &  $0.59 \pm 0.01^{\rm (b)}$ & $0.61 \pm 0.02^{\rm (c)}$  & $0.57 \pm 0.02^{\rm (d)}$ &  $0.69 \pm 0.04^{\rm (e)}$\\ 
By comparing $\overline{\Delta  \Pi}$ with $\Delta \Pi_{\ell}^{\rm a}$ (this work) &  $0.575 \pm 0.005$  & $0.615 \pm 0.015$  & $0.595 \pm 0.015$  & $0.70 \pm 0.02$ \\ 
By comparing $\overline{\Delta  \Pi}$ with $\overline{\Delta \Pi_{k}}$ (this work) &  $0.555  \pm 0.005$  & $0.575  \pm 0.015$ & $0.565 \pm 0.015$  & $0.64 \pm 0.02$ \\ 
By means of $\chi^2 $ period fittings (this work)     &  $0.56 \pm 0.01$ & $0.60 \pm 0.01$  & $0.55 \pm 0.01$  &  $0.64 \pm 0.01$  \\
\hline
\hline
\end{tabular}

{\footnotesize References: 
(a) Dreizler \& Heber (1998); 
(b) KB94; 
(c) Reed et al. (2000); 
(d) Kawaler et al. (2004); 
(e) O'Brien (2000)}
\end{table*}

In  order to  infer the  structural parameters  of these  four  GW Vir
stars, we shall employ three different methods. First, we estimate the
stellar masses by  using the asymptotic period spacing  of our models,
as  computed  from Eq.  (\ref{asymptotic}).  Specifically,  we make  a
direct comparison between $\Delta \Pi_{\ell}^{\rm a}$ and the observed
mean  period  spacing   $\overline{\Delta  \Pi}$,  assuming  that  the
effective temperature of the target  star is that obtained by means of
spectroscopy.  In  the second approach,  we repeat this  procedure but
this  time   using  the  average  of  the   computed  period  spacings
$\overline{\Delta \Pi_{k}}$  and comparing  it with the  observed mean
period  spacing  $\overline{\Delta \Pi}$.   The  third  approach is  a
fitting method  in which we compare the  theoretical ($\Pi_k^{\rm T}$)
and observed ($\Pi_i^{\rm O}$) periods by means of a standard $\chi^2$
algorithm. In the three approachs  we assume that the observed periods
correspond to $\ell=1, m=0$ modes.

\subsection{Inferences from the asymptotic period spacing, 
$\Delta \Pi_{\ell}^{\rm a}$}
\label{inf-asym}

\begin{figure}
\centering
\includegraphics[clip,width=250pt]{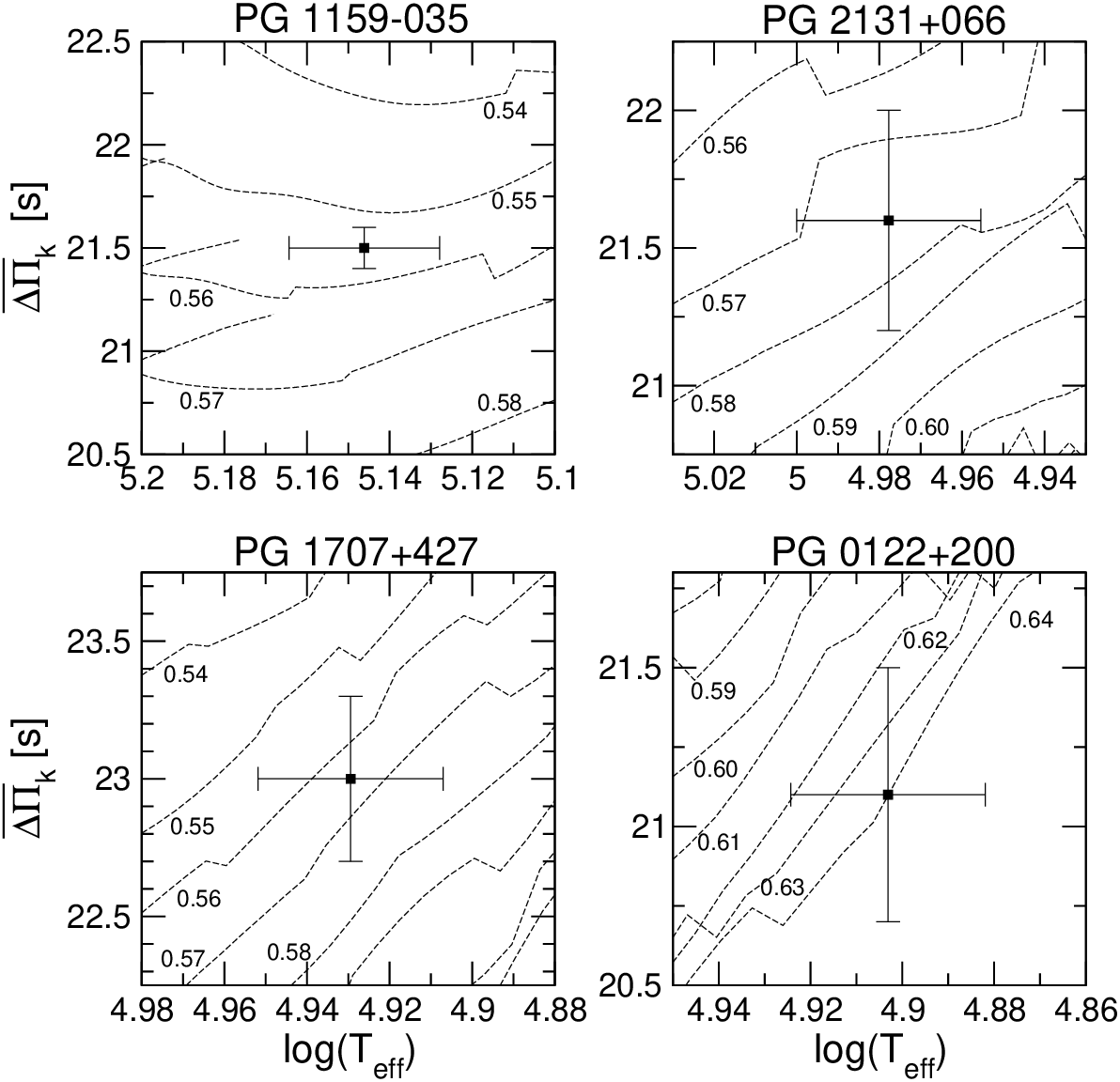}
\caption{The average of the computed period spacings 
for  PG 1159  models with  different stellar  masses in  terms  of the
effective  temperature. Each panel  corresponds to  a specific  GW Vir
star.   The observed  mean period  spacing $\overline{\Delta  \Pi}$ is
also shown (taken from Kawaler et al. 2004).}
\label{acps}
\end{figure}

We  start by  examining  Fig. \ref{asint2},  in  which the  asymptotic
period spacing ($\Delta \Pi_{\ell}^{\rm a}$)  is shown in terms of the
effective  temperature for PG  1159 sequences  with stellar  masses of
$0.53,\cdots,0.64 M_{\odot}$. We also include in the figure the values
of  the   observed  mean  period   spacing,  $\overline{\Delta  \Pi}$,
corresponding  to  the GW  Vir  stars  PG  1159-035, PG  2131+066,  PG
1707+427, and  PG 0122+200. The values of  $\overline{\Delta \Pi}$ and
the associated  error bars are  taken from Table  6 of Kawaler  et al.
(2004).   In  Table  1  we  summarize  our  results.   Note  that  the
spectroscopic estimation of $T_{\rm eff}$  for PG 0122+200 place it at
an effective temperature that is  lower than our PG 1159 models. Thus,
for  PG 0122+200 we  can only  make a  rough extrapolation.   Note the
notable  agreement between our  estimates and  the values  inferred by
other  pulsation  studies.   Note  also that  the  seismic  inferences
suggest  higher  mass  values   as  compared  with  the  spectroscopic
estimations.  We  do  not  actually  understand  the  origin  of  this
discrepancy, but we note that spectroscopic derivations of the stellar
mass are  usually very uncertain due  to the large  uncertainty in the
determination of $\log g$ --- an error of $0.3$ dex translates into an
error of  $0.1 M_{\odot}$ (Dreizler \& Heber  1998).  Indeed, Dreizler
\& Heber  (1998) first estimate $\log  g$ and $T_{\rm  eff}$ values by
using fits to model atmospheres, and then they select the stellar mass
from the evolutionary  tracks of O'Brien (2000). Thus,  both the large
uncertainties in the estimation of $\log g$ ($\approx$ 0.5 dex) and in
the evolutionary  computations could account for  an excessively broad
range of allowed masses for a specific star.
 
\subsection{Inferences from the average  of the computed 
period spacings, $\overline{\Delta \Pi_{k}}$}
\label{inf-aver}

Here, we repeat the procedure described in \S \ref{inf-asym}, but this
time  by  employing  the  average  of  the  computed  period  spacings
($\overline{\Delta \Pi_{k}}$)  instead the asymptotic  period spacing.
We   compare    $\overline{\Delta   \Pi_{k}}$   with    the   measured
$\overline{\Delta  \Pi}$  for   each  star  under  consideration.   In
computing  the average of  the computed  period spacings,  we consider
only the  period interval in which  the periodicities of  a given star
are observed. For instance, for PG 1159-035, we compute the average of
the  computed period  spacings for  periods  in the  range [430,  841]
s. The  results of  our calculations are  shown in  Fig.\ref{acps}, in
which we show $\overline{\Delta \Pi_{k}}$ plotted with dashed lines in
terms   of  $T_{\rm   eff}$  for   the   four  GW   Vir  stars   under
consideration. Note that,  since we have performed the  average of the
computed period  spacings on  different period ranges,  appropriate to
the  period  range  exhibited  by  a  specific  star,  the  curves  of
$\overline{\Delta
\Pi_{k}}$  are different in  each panel.   In Table  1 we  include the
estimations of  the stellar  mass for the  four stars.  Note  that the
stellar masses  as determined by  this approach are  appreciably lower
than the  values inferred by  using the asymptotic period  spacing and
thus in  better agreement with spectroscopic inferences.   This is due
to  $\overline{\Delta  \Pi_{k}}$  is  typically  $\approx  0.8-1.0$  s
smaller than $\Delta \Pi_{\ell}^{\rm  a}$, irrespective of the stellar
mass and effective temperature.  Thus, if we compare the observed mean
period spacing for a given star with $\overline{\Delta
\Pi_{k}}$ we obtain a smaller total mass ($0.02-0.06 M_{\odot}$ lower)
than that inferred by comparing  the observed mean period spacing with
$\Delta  \Pi_{\ell}^{\rm a}$.   We stress  that the  asymptotic period
spacing  $\Delta \Pi_{\ell}^{\rm  a}$,  as computed  by  means of  Eq.
\ref{asymptotic}, is formally valid for the limit of high radial order
$k$  in chemically homogeneous  stars. Because  our PG1159  models are
chemically stratified, we conclude  that the estimations of $M_*$ from
$\overline{\Delta \Pi_{k}}$ are more  realistic than those inferred by
means of  $\Delta \Pi_{\ell}^{\rm a}$.   We also found that  our $M_*$
values are lower than those quoted by other seismic studies (see Table
1).  We note  that, with  the exception  of the  work of  KB94  for PG
1159-035, all  these studies compare the observed  mean period spacing
with the asymptotic  predictions to infer the stellar  mass. Thus, not
surprisingly, their  values are very  similar to our  results obtained
from the asymptotic period spacing in
\S \ref{inf-asym},  but rather departed  from our values  deduced from
the average of the computed period spacings.

\begin{figure}
\centering
\includegraphics[clip,width=250pt]{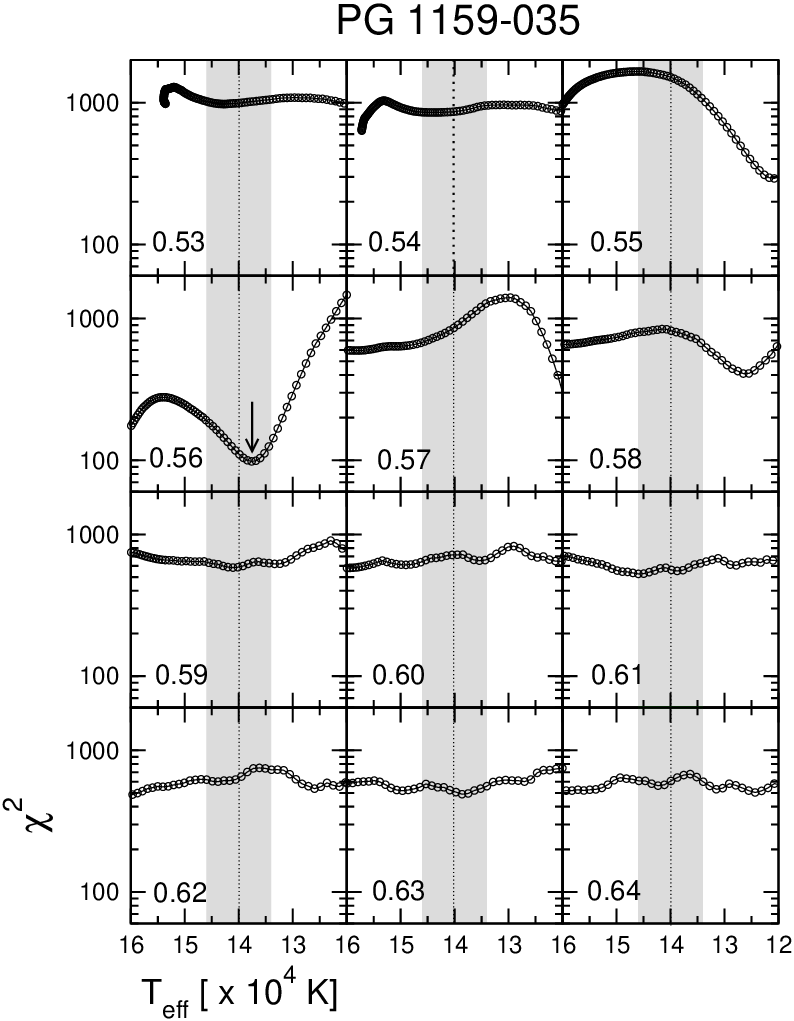}
\caption{The $\chi^2$ values as defined by Eq. (\ref{chi2}) 
corresponding  to PG  1159-035  versus the  effective temperature  for
$\ell=1$ modes.   Each panel depicts  the results for a  specific mass
value. The  gray strip  corresponds to the  allowed values  of $T_{\rm
eff}$ according to spectroscopic  measurements. The arrow indicate the
minimum value of $\chi^2$ associated with the effective temperature of
the best-fit model.}
\label{chipg1159}
\end{figure}

\subsection{Inferences from the computed pulsation periods,
$\Pi_k$}

We turn now to the $\chi^2$ period fitting procedure. The goodness of 
a given  fit is measured by means of a quality function, defined as

\begin{equation}
\chi^2(M_*, T_{\rm eff})=
\sum_{i=1}^{N} \min[(\Pi_i^{\rm O}- \Pi_k^{\rm T})^2]. 
\label{chi2}
\end{equation}

\begin{table}
\centering
\caption{Comparison between the observed periods 
of PG 1159-035 (taken from KB94) 
and theoretical ($\ell=1$) periods of the 
best-fit model ($\overline{\delta \Pi}= 1.79$ s, $\sigma_{\delta \Pi}= 2.21$ s).}
\begin{tabular}{cccrr}
\hline
\hline
$\Pi_i^{\rm O}$ & $\Pi_k^{\rm T}$ & $k$ & 
$\delta \Pi_i$ & $\delta \Pi_i/ \Pi_i$\\
 $[$s$]$ & $[$s$]$ & & $[$s$]$  & $[\%]$ \\
\hline
$ 430.04 $ & $ 433.21 $ & 18 & $ -3.17 $ & $ -0.74 $ \\ 
$ 450.71 $ & $ 453.49 $ & 19 & $ -2.78 $ & $ -0.62 $ \\ 
$ 469.57 $ & $ 473.51 $ & 20 & $ -3.94 $ & $ -0.84 $ \\ 
$ 494.85 $ & $ 496.58 $ & 21 & $ -1.73 $ & $ -0.35 $ \\ 
$ 517.18 $ & $ 517.10 $ & 22 & $  0.08 $ & $  0.02 $ \\ 
$ 538.16 $ & $ 537.50 $ & 23 & $  0.66 $ & $  0.12 $ \\ 
$ 558.44 $ & $ 559.15 $ & 24 & $ -0.71 $ & $ -0.13 $ \\ 
$ 581.29 $ & $ 580.55 $ & 25 & $  0.74 $ & $  0.13 $ \\ 
$ 603.04 $ & $ 601.28 $ & 26 & $  1.76 $ & $  0.29 $ \\ 
$ 622.60 $ & $ 622.60 $ & 27 & $  0.00 $ & $  0.00 $ \\ 
$ 643.41 $ & $ 645.24 $ & 28 & $ -1.83 $ & $ -0.28 $ \\ 
$ 666.22 $ & $ 665.09 $ & 29 & $  1.13 $ & $  0.17 $ \\ 
$ 687.71 $ & $ 686.29 $ & 30 & $  1.42 $ & $  0.21 $ \\ 
$ 707.92 $ & $ 709.95 $ & 31 & $ -2.03 $ & $ -0.29 $ \\ 
$ 729.50 $ & $ 729.65 $ & 32 & $ -0.15 $ & $ -0.02 $ \\ 
$ 753.12 $ & $ 750.01 $ & 33 & $  3.11 $ & $  0.41 $ \\ 
$ 773.77 $ & $ 773.33 $ & 34 & $  0.44 $ & $  0.06 $ \\ 
$ 790.94 $ & $ 795.17 $ & 35 & $ -4.23 $ & $ -0.53 $ \\ 
$ 817.12 $ & $ 814.96 $ & 36 & $  2.16 $ & $  0.26 $ \\ 
$ 840.02 $ & $ 836.27 $ & 37 & $  3.75 $ & $  0.45 $ \\
\hline
\hline
\end{tabular}
\end{table}

\noindent Here, $\Pi_i^{\rm O}$ is one of the $N$ observed pulsation 
periods, and $\Pi_k^{\rm T}$ is a specific computed period with radial
index $k$.  The method consists of  looking for the PG 1159 model that
shows the lowest value of $\chi^2$.  Obviously, $\chi^2 \rightarrow 0$
if the match  between observed and computed periods  were perfect.  We
evaluate  $\chi^2(M_*, T_{\rm  eff})$ varying  $M_*/M_{\odot}$  in the
range  $[0.53,0.64]$  with  a  step  $\Delta  (M_*/M_{\odot})=  0.01$,
whereas for the effective temperature we adopt a much more finer grid,
with $\Delta  T_{\rm eff}=  10-20$ K.

\begin{figure}
\centering
\includegraphics[clip,width=250pt]{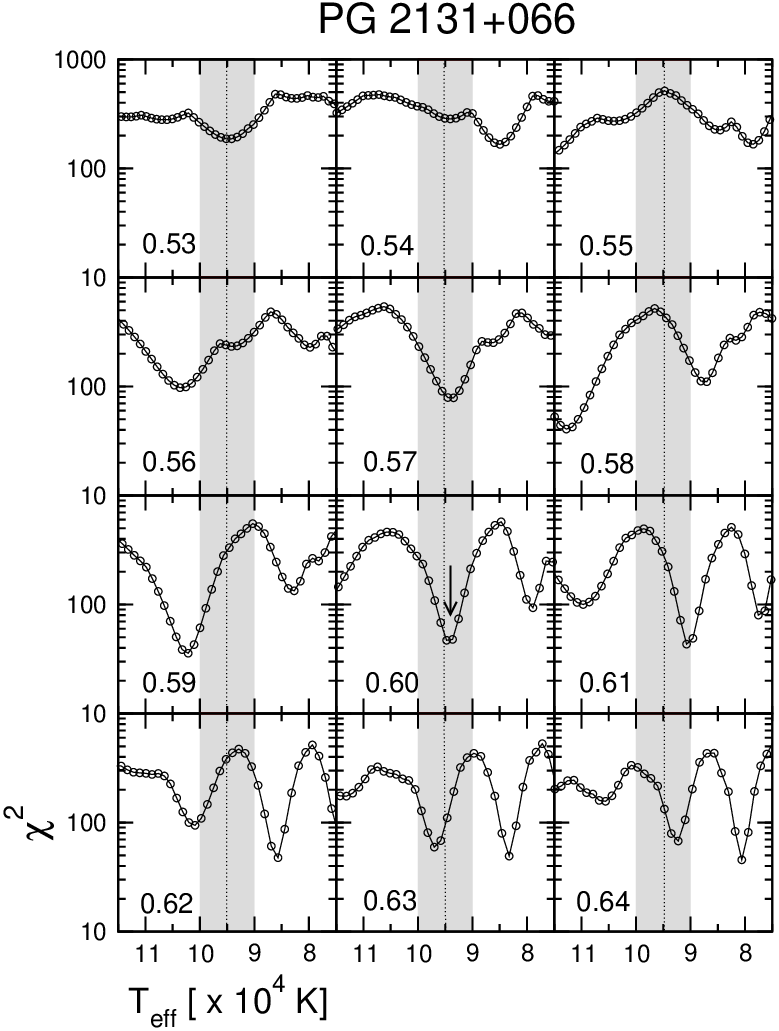}
\caption{Same as Fig. \ref{chipg1159}, but for PG 2131+066.}
\label{chipg2131}
\end{figure}

We  initially searched  the optimal  model within  a wide  interval of
effective temperatures, ranging from  the hottest point reached by the
model  in the  H-R diagram  ($160\,000-200\,000$ K,  according  to the
value  of  $M_*$)  until  $T_{\rm  eff}  \sim  70\,000$  K.   However,
throughout  our calculations  we realized  that generally  the  run of
$\chi^2$ as a function of $T_{\rm eff}$ for a given star exhibits more
than one minimum,  meaning that the period spectrum  of the star could
not  be fitted  by an  unique PG  1159 model.  This behaviour  is more
pronounced in the  case of the GW Vir stars  with few observed periods
available.   This effect  can  be  understood on  the  basis that  the
pulsation periods  of a specific  model are generally  increasing with
time. If at a given effective  temperature the model shows a close fit
to the  observed periods, then  the function $\chi^2$ reaches  a local
minima.  When the model has cooled  enough (at some time later), it is
possible that  the accumulated period drift nearly  matches the period
spacing   between  adjacent   modes  ($|\Delta   k|=  1$).   In  these
circumstances, the model is able to fit the star again, as a result of
which  $\chi^2$   exhibits  other   local  minimum.   To   break  this
degeneracy,  we  decided  to  consider  a  more  restricted  range  of
effective  temperatures,  but  that  still  comfortably  embraces  the
spectroscopic  estimation  of  $T_{\rm  eff}$ and  its  uncertainties.
Specifically, in searching the  ``best-fit'' model for a specific star
we considered effective temperatures laying inside an interval $T_{\rm
eff}^{\rm spec}- 20,000 \leq  T_{\rm eff} \leq T_{\rm eff}^{\rm spec}+
20\,000$  K ,  where  $T_{\rm eff}^{\rm  spec}$  is the  spectroscopic
determination  of  the effective  temperature.  As  we shall  describe
below, for the star PG 1159-035  we were able to find a best-fit model
for which  $\chi^2$ adopts  the lowest value  within this  interval of
$T_{\rm eff}$s.  The situation  was considerably more difficult for PG
2131+066,  PG  1707+427, and  PG  0122+200,  due  to the  scarcity  of
pulsation periods characterizing their pulsation spectra. In fact, for
these stars  the ambiguity associated  with the existence  of multiple
solutions  remained even  when we  adopted a  more  restricted $T_{\rm
eff}$-interval, and in that cases  the procedure failed to isolate the
best solution. In  order to solve this situation,  for PG 2131+066, PG
1707+427,  and PG  0122+200  we  restricted the  search  to within  $1
\sigma$ of the spectroscopic value.

\begin{table}
\centering
\caption{Comparison between the observed periods 
of PG 2131+066 (taken from Kawaler et al. 1995) 
and the theoretical ($\ell=1$) periods of the 
best-fit model ($\overline{\delta \Pi}= 1.83$ s, 
$\sigma_{\delta \Pi}= 2.59$ s).}
\begin{tabular}{cccrr}
\hline
\hline
$\Pi_i^{\rm O}$ & $\Pi_k^{\rm T}$ & $k$ & $\delta \Pi_i$ & 
$\delta \Pi_i/ \Pi_i$\\
 $[$s$]$ & $[$s$]$ & & $[$s$]$  & $[\%]$ \\
\hline
$ 341.45 $ & $ 341.40 $ & 14 & $ 0.05  $ & $ 0.014 $ \\
---        & $ 364.21 $ & 15 & ---       & ---      \\
$ 384.27 $ & $ 382.23 $ & 16 & $ 2.04  $ & $ 0.53  $ \\
$ 403.93 $ & $ 403.86 $ & 17 & $ 0.07  $ & $ 0.017 $ \\
$ 426.36 $ & $ 425.07 $ & 18 & $ 1.29  $ & $ 0.30  $ \\
$ 450.28 $ & $ 445.17 $ & 19 & $ 5.11  $ & $ 1.13  $ \\
$ 462.39 $ & $ 466.26 $ & 20 & $ -3.87 $ & $ -0.84 $ \\
---        & $ 486.74 $ & 21 & ---      & ---      \\
$ 507.91 $ & $ 508.27 $ & 22 & $ -0.36 $ & $ -0.07 $ \\
\hline
\hline
\end{tabular}
\end{table}

We  start by  examining  Fig.   \ref{chipg1159} in  which  the run  of
$\chi^2$ is plotted versus $T_{\rm  eff}$ for PG 1159-035.  We include
the results for the complete  set of stellar masses considered in this
work.   The  observed periods  are  the  same  twenty $\ell=  1,  m=0$
consecutive  periods considered  in  KB94.  The  vertical dotted  line
corresponds to the effective temperature of PG 1159-035 as measured by
means of spectroscopic  techniques ($T_{\rm eff}^{\rm spec}= 140\,000$
K). Note  that there is  a well-defined minima of  $\chi^2$, indicated
with an arrow in the plot, corresponding to a stellar model with $M_*=
0.56 M_{\odot}$ and $T_{\rm eff} \sim 137,000$ K.  We adopt this model
as our best-fit  model.  A comparison between the  observed periods of
PG 1159-035 and the  theoretical periods corresponding to the best-fit
model is  included in  Table 2.  The  first column lists  the observed
periods, and the  second and third columns correspond  to the computed
periods and  the associated radial order, respectively.   The last two
columns show  the difference and the relative  difference (in percent)
between the  observed and  computed periods, respectively,  defined as
$\delta  \Pi_i= \Pi_i^{\rm  O}  - \Pi_k^{\rm  T}$  and $\delta  \Pi_i/
\Pi_i= 100 \times (\Pi_i^{\rm O} -
\Pi_k^{\rm T}) / \Pi_i^{\rm  O}$. Note  that the period  match 
is excellent.   To quantitatively  measure the quality  of the  fit we
compute the  mean of  the period differences,  $\overline{\delta \Pi}$
and the  root-mean-square residual, $\sigma_{\delta  \Pi}$.  We obtain
$\overline{\delta \Pi}= 1.79$ s  and $\sigma_{\delta \Pi}= 2.21$ s for
the fit to  PG 1159-035. The quality of our fit  is comparable to that
achieved by  KB94 ($\overline{\delta \Pi} \sim 1.19$  s), although the
KB94 fit is notoriously better.  In this connection, it is interesting
to  note  that  these  authors  adopt $M_{\rm  He}$  and  the  surface
composition  as free  adjustable  parameters for  the  period fit,  in
addition  to   the  stellar   mass  and  the   effective  temperature.
Therefore,  they are  able to  make a  finner tuning  of  the observed
period  spectrum  than  our  own.  The structural  parameters  of  our
best-fit model are listed in  Table 6.  Note the rather nice agreement
between our  predictions and the structural properties  of PG 1159-035
as inferred by other standard techniques.  For instance, the effective
temperature, the  luminosity and the  surface gravity of  our best-fit
model are  consistent at the  $1 \sigma$ level with  the spectroscopic
estimations of  Dreizler \& Heber  (1998). In particular,  the quality
function of our period fit naturally adopts an unique minimum value at
an effective  temperature very close to the  spectroscopic value (this
is not the case for the remainder GW Vir stars; see below).  Note also
that the  total mass of our  model lies at $0.02  M_{\odot}$ above the
spectroscopic mass, although well within the allowed range ($0.44 \leq
M_*/M_{\odot}  \leq 0.64$).  We stress  again that  spectroscopy
determines only $\log g$ and $T_{\rm  eff}$, and then the value of the
stellar mass  is assessed  using  evolutionary tracks, which  depend on
the  complete  evolution  of  the  progenitors.   We  note  that  our
seismological  mass  is lower  than  the  value  obtained by  KB94  by
employing the same set of observed pulsation periods.

\begin{table}
\centering
\caption{Comparison between the observed periods 
of PG 1707+427 (taken from Kawaler et al. 2004) and 
the theoretical ($\ell=1$) periods of the best-fit model 
($\overline{\delta \Pi}= 2.31$ s, $\sigma_{\delta \Pi}= 2.70$ s).}
\begin{tabular}{cccrr}
\hline
\hline
$\Pi_i^{\rm O}$ & $\Pi_k^{\rm T}$ & $k$ & $\delta \Pi_i$ & 
$\delta \Pi_i/ \Pi_i$\\
 $[$s$]$ & $[$s$]$ & & $[$s$]$  & $[\%]$ \\
\hline
$ 334.62 $ & $  332.14 $ & 12 & $ 2.48 $ & $0.74$ \\
  ---      & $  352.76 $ & 13 &  ---     & ---    \\
  ---      & $  377.17 $ & 14 &  ---     & ---    \\
  ---      & $  399.73 $ & 15 &  ---     & ---    \\
  ---      & $  420.53 $ & 16 &  ---     & ---    \\
$ 448.07 $ & $  445.72 $ & 17 & $ 2.35 $ & $0.52$ \\
  ---      & $  467.50 $ & 18 &  ---     & ---    \\
$ 494.39 $ & $  489.62 $ & 19 & $ 4.77 $ & $0.96$ \\
  ---      & $  514.23 $ & 20 &  ---     & ---    \\
$ 536.41 $ & $  535.84 $ & 21 & $ 0.57 $ & $0.11$ \\
  ---      & $  559.94 $ & 22 &  ---     & ---    \\
  ---      & $  581.88 $ & 23 &  ---     & ---    \\
  ---      & $  606.38 $ & 24 &  ---     & ---    \\
  ---      & $  628.83 $ & 25 &  ---     & ---    \\
  ---      & $  651.41 $ & 26 &  ---     & ---    \\
$ 677.89 $ & $  677.48 $ & 27 & $ 0.41 $ & $0.06$ \\
  ---      & $  697.36 $ & 28 &  ---     & ---    \\
$ 726.02 $ & $  723.05 $ & 29 & $ 2.97 $ & $0.41$ \\
$ 745.78 $ & $  747.18 $ & 30 & $ - 1.4$ & $-0.19$ \\
  ---      & $  769.01 $ & 31 &  ---     & ---    \\
  ---      & $  792.73 $ & 32 &  ---     & ---    \\
  ---      & $  818.54 $ & 33 &  ---     & ---    \\ 
  ---      & $  841.27 $ & 34 &  ---     & ---    \\
  ---      & $  862.57 $ & 35 &  ---     & ---    \\
  ---      & $  890.30 $ & 36 &  ---     & ---    \\
$ 909.05 $ & $  912.62 $ & 37 & $-3.57 $ & $-0.39$ \\
\hline
\hline
\end{tabular}
\end{table}

\begin{table}
\centering
\caption{Comparison between the observed periods 
of PG 0122+200 (taken from Vauclair et al. 2001) and the
theoretical ($\ell=1$) periods of the best-fit model 
($\overline{\delta \Pi}= 2.61$ s, $\sigma_{\delta \Pi}= 3.55$ s).}
\begin{tabular}{cccrr}
\hline
\hline
$\Pi_i^{\rm O}$ & $\Pi_k^{\rm T}$ & $k$ & $\delta \Pi_i$ & 
$\delta \Pi_i/ \Pi_i$\\
 $[$s$]$ & $[$s$]$ & & $[$s$]$  & $[\%]$ \\
\hline
$336.67$ & $340.30$ & 14 & $-3.63$  & $-1.08$   \\
---      & $360.88$ & 15 & --- & ---  \\
$379.60$ & $381.16$ & 16 & $-1.56$  & $-0.41$   \\
$401.00$  & $401.75$ & 17 & $-0.75$  & $-0.19$   \\
---      & $424.05$ & 18 & --- & ---  \\
$449.43$ & $442.62$ & 19 & $6.81$  & $1.51$   \\
$468.69$ & $463.65$ & 20 & $5.04$  & $1.08$  \\
---      & $485.45$ & 21 & --- & ---  \\
---      & $506.63$ & 22 & --- & ---  \\
---      & $527.22$ & 23 & --- & ---  \\
---      & $546.99$ & 24 & --- & ---  \\
$570.00$ & $570.35$ & 25 & $-0.35$  & $-0.06$   \\
---      & $591.06$ & 26 & --- & ---  \\
$611.05$ & $611.24$ & 27 & $-0.19$  & $-0.03$   \\
\hline
\hline
\end{tabular}
\end{table}

At variance with  PG 1159-035, the stars PG  2131+066, PG 1707+427 and
PG  0122+200  exhibit  very  few  pulsation  periods  in  their  light
curves. As  we already anticipated, this unfortunate  fact renders any
asteroseismic  inference on these  stars considerably  more difficult.
This is the reason for  which most of asteroseismic studies carried so
far rely on the measured period  spacing only.  In spite of this fact,
we have attempted to apply  the $\chi^2$ fitting method to these stars
by adopting a more strict criteria, i.e., by considering as valid only
solutions  that guaranteed consistency  at $1  \sigma$ level  with the
spectroscopic  determination of  $T_{\rm eff}$.  Fig.  \ref{chipg2131}
shows the  results for  PG 2131+066. For  this star  the observational
data consists of only the  seven pulsation periods reported by Kawaler
et al. (1995).   Note that, at variance with the  case of PG 1159-035,
the  function $\chi^2$  shows various  minima compatible  with $T_{\rm
eff}^{\rm spec}$  in the  range of effective  temperatures considered.
For instance, for  $M_*= 0.60 M_{\odot}$ there is  a remarkable minima
in the immediate vicinity  of the spectroscopic estimation, at $T_{\rm
eff}  \approx 95\,000$ K.   Other comparable  minimum of  $\chi^2$ are
seen also for stellar masses of $0.61, 0.63$ and $0.64 M_{\odot}$, but
at  effective temperatures  somewhat departed  from  the spectroscopic
value.  These minima represent  equivalent solutions from the point of
view  of  the magnitude  of  $\chi^2$,  and  thus, there  are  various
possible optimal  models. We elect to  adopt as the  best-fit model to
that  having the  effective  temperature closer  to the  spectroscopic
determination,  that is,  the  model with  $M_*=  0.60 M_{\odot}$  and
$T_{\rm   eff}=    94\,740$   K   (indicated   with    an   arrow   in
Fig. \ref{chipg2131}).   A comparison between the  observed periods of
PG 2131+066  and the theoretical  ($\ell= 1$) periods of  the best-fit
model is given in Table 3. We note that the overall quality of the fit
($\overline{\delta \Pi}=  1.83$ s,  $\sigma_{\delta \Pi}= 2.59$  s) is
nearly so satisfactory  as in the case of PG  1159-035 (Table 2).  The
characteristics of  our best-fit model  for PG 2131+066 are  listed in
Table  6.   Our  results  are  in good  agreement  with  spectroscopy,
although our determination of the stellar mass is significantly higher
--- but  even consistent  at  the  $1 \sigma$  level.   Note that  the
disagreement  in the  estimation of  the  stellar mass  would be  more
pronounced  if  we  were  adopting  anyone  of  the  other  acceptable
solutions, for which $M_*
\gtrsim 0.60 M_{\odot}$.

\begin{table*}
\centering
\caption{Summary of the seismic inferences for PG 1159-035, 
PG 2131+066,  PG 1707+427 and  PG 0122+200 from $\chi^2$  period fits.
In  the  interest  of  comparison,  we include  values  of  structural
properties extracted from  spectroscopic determinations and the values
of the observed mean period  spacing extracted from Table 6 of Kawaler
et al. (2004).}
\begin{tabular}{l|cc|cc|cc|cc}
\hline
\hline
Quantity & Model & PG 1159-035 & Model & PG 2131+066 & Model & PG 1707+427 & Model  & PG 0122+200 \\
\hline
$\overline{\Delta \Pi}$ [s]         & 21.21   & $21.5 \pm 0.1$   & $20.86$ & $21.6 \pm 0.4$       & 23.22   & $23.0  \pm 0.3$      & 20.92  & $21.10 \pm 0.4$   \\
$T_{\rm eff}$ [K]                   & 137\,620  & $140\,000 \pm 6\,000$   & 94\,740   & $95\,000 \pm 5\,000$      & 87\,585   & $85\,000 \pm 4\,500$     & 81\,810  & $80\,000 \pm 4\,000$  \\ 
$M_*/M_{\odot}$                     & 0.56   & $0.54 \pm 0.1$       & 0.60   & $0.55 \pm 0.1$         & 0.55    & $0.54  \pm 0.1$      & 
0.64   & $0.54  \pm 0.1$   \\
$\log g$ [cm/s$^2$]                 & 7.31    & $7.0 \pm 0.3$       & 7.71    & $7.5 \pm 0.3$         & 7.59    & $7.5   \pm 0.3$      &  7.89      & $7.5   \pm 0.3$   \\
$\log (L_*/L_{\odot})$                & 2.38    & $2.7 \pm 0.4$       & 1.37    & $1.6 \pm 0.3$         & 1.31    & $1.4   \pm 0.3$      & 0.96   & $1.3   \pm 0.3$   \\  
$R_*/R_{\odot}$                     & 0.0273  & ---                 & 0.0179  & ---                      & 0.02    & ---                  &  0.015      & ---               \\
$M_{\rm He}$ $[10^{-3} M_{\odot}]$  & $4.94$  & ---                 & $5.29$  & ---                      & $4.85$  & ---                  & $5.65$ & ---               \\   
distance [pc]                       & $482 \pm 44$         &  $800^{+ 600 {\rm (a)}}_{- 400}$      & $716^{+ 185}_{-147}$ &  $681^{+ 170 {\rm (b)}}_{-137}$    &   $697^{+ 180}_{-144}$       & $1300^{+ 1000 {\rm (a)}}_{-600}$        &        &                   \\ 
\hline
\hline
\end{tabular}

{\footnotesize References: 
(a) Werner et al. (1991);
(b) Reed et al. (2000)} 
\end{table*}

We have  also applied  our fitting  procedure to the  GW Vir  stars PG
1707+427 and PG 0122+200. We  have found that the behavior of $\chi^2$
in these  cases is,  not surprisingly, very  similar to that  shown by
Fig.   \ref{chipg2131}  for PG  02131+066,  because  these stars  also
exhibit  a very reduced  number of  pulsation periods.   Therefore, to
obtain optimal representative models we employ the same procedure than
in the case of PG 02131+066,  i.e., we consider as valid the solutions
that are consistent at  $1 \sigma$ with the spectroscopically inferred
effective temperature. In this way we discard other possible solutions
at  effective  temperatures  that   do  not  match  the  spectroscopic
evidence.  The results are summarized in Tables 4 and 5, respectively.
Note that  for PG  1707+427 the  period match is  of a  slightly lower
quality than for PG 1159-035  and PG 2131+066, but still satisfactory,
with $\overline{\delta  \Pi}= 2.31$ s and  $\sigma_{\delta \Pi}= 2.70$
s.  For  PG 0122+200,  however, the fit  is considerably  worse, being
$\overline{\delta  \Pi}= 2.61$  s and  $\sigma_{\delta \Pi}=  3.55$ s.
The main source of discrepancy  comes from the existence of periods at
449.43 and  468.69 s.  We mention  that for PG  0122+200 we  have also
considered  the  set  of  observed  periods  reported  by  O'Brien  et
al. (1998), which do not includes the period at 468.69 s. We find that
the quality  of the  fit in that  case do not  improves significantly.
The structural properties  of the best-fit models for  PG 1707+427 and
PG 0122+200 are  given in Table 6. For PG 1707+427  we found a general
agreement  between our  inferences  and the  spectroscopic values,  in
particular concerning the stellar mass.  In the case of PG 0122+200 we
found a total  mass and a surface gravity large  in excess as compared
with the spectroscopic evidence. This  high mass value is in line with
other seismic  determinations and also with our  own predictions based
on  the asymptotic  period spacing  and  the average  of the  computed
period spacings (see Table1).

In addition  to the  structural properties of  the GW Vir  stars under
study, we infer  their ``seismic'' distance from the  Earth. First, we
compute the  bolometric magnitude from the luminosity  of the best-fit
model, by means of $M_{\rm bol}= M_{\odot {\rm bol}} - 2.5
\log(L_*/L_{\odot})$,   with  $M_{\odot   {\rm  bol}}=   4.75$  (Allen
1973). Next,  we transform the bolometric magnitude  into the absolute
magnitude, $M_{\rm  v}= M_{\rm bol} -  {\rm BC}$, where  ${\rm BC}$ is
the bolometric  correction. Finally,  we compute the  seismic distance
according to $\log d\ [{\rm pc}]= \frac{1}{5}
\left( m_{\rm v} - M_{\rm v} +5 \right)$.  
For PG 1159-035 ($m_{\rm v}=  14.84$) we adopt a bolometric correction
of  ${\rm  BC}=  -7.6  \pm  0.2$  from  KB94.   Following  Kawaler  et
al. (1995), we adopt ${\rm BC}= -6.0 \pm 0.5$ for PG 2131+066 ($m_{\rm
v}= 16.6$) and PG 1707+427 ($m_{\rm v}= 16.69$).  Unfortunately, there
is not  available any estimation  of the bolometric correction  for PG
0112+200, hindering  any attempt to  infer its seismic  distance.  Our
seismic distances  are shown in Table  6. In addition,  we include the
distances  estimated by  means of  other techniques.  We found  a good
agreement  (within $1  \sigma$) between  our distances  and  the other
non-seismic  estimations,  although our  values  are characterized  by
errors  much  smaller.  Our  results  are  also  consistent  with  the
distances obtained  seismologically by KB94  for PG 1159-035  ($d= 440
\pm  40$  pc)  and  by  Reed  et al.   (2000)  for  PG  2131+066  ($d=
668^{+78}_{-83}$ pc).

\section{Conclusions} 

In this paper  we have studied in detail some  relevant aspects of the
adiabatic  pulsations  of  GW  Vir  stars  by  using  state-of-the-art
evolutionary  PG 1159  models  recently presented  by  Althaus et  al.
(2005).   As far as  we are  aware, this  is the  first time  that the
adiabatic pulsation  properties of fully  evolutionary post born-again
PG 1159  models like the ones  employed in this work  are assessed. We
refer the interested reader to the paper by Gautschy et al. (2005) for
details about the non-adiabatic pulsation properties of these models.

We first explored the basic nature of the pulsation modes by employing
propagation  diagrams and  weight  functions.  In  line with  previous
works, our results suggest that  for PG 1159 models at high luminosity
stages,  the propagation diagrams  are reminiscent  of those  of their
progenitors --- red  giant stars --- characterized by  large values of
the Brunt-V\"ais\"al\"a frequency in  the central regions of the star.
As a  result, pulsation $g$-modes ---  the only observed so  far in GW
Vir  stars ---  are strongly  confined to  the highly  condensed core,
whereas $p$-modes are  free to oscillate in more  external regions. In
addition, we  have found that  there exist several modes  exhibiting a
mixed  nature,  behaving  as  $p$-modes  in  outer  regions  and  like
$g$-modes  in the  deeper zones  within  the star.   As the  effective
temperature  increases  these  modes  undergoes  several  episodes  of
avoided crossing.  We found  that pulsation periods generally decrease
with time, an effect attributable to the rapid contraction experienced
by the star  in their incursion to the blue in  the H-R diagram.  Note
that  a decreasing  period implies  negative temporal  rate  of period
change; according to  our results, this is the case  for all the modes
during  the  contraction stage.  Once  the  models  have passed  their
maximum  effective temperature and  have settled  onto the  WD cooling
track,  the  Brunt-V\"ais\"al\"a frequency  acquires  a more  familiar
shape, typical of  the WD pulsators.  In this  phase all the pulsation
periods increase  with decreasing effective  temperature, reflecting a
lowering in the magnitude  of the Brunt-V\"ais\"al\"a frequency at the
core regions. At this stages, $g$-modes become envelope modes, and, as
indicated by the  weight functions, the outer regions  of the star are
the more relevant ones in  establishing the periods.  By the contrary,
pulsation $p$-modes  are almost confined  to the degenerate  core.  We
note that, since all the  periods increase during this phase, the rate
of  period  change  for  any  pulsation  mode  must  be  positive.  In
particular, the measured rate of  period change of the period at $\sim
516$ s  in PG 1159-035 is  positive (Costa et al.  1999), in agreement
with our findings. We defer to  a future work a complete discussion of
the rate of period change of our PG 1159 models.

We next have focused our attention on the evolutionary stages in which
the star has already entered its WD cooling track. For these phases we
have  been  able  to  obtain additional  evolutionary  sequences  with
several values of  the stellar mass, allowing us  to study the effects
of $M_*$ and the effective  temperature on the pulsation properties of
our models. In particular, we have examined the asymptotic behavior of
the $g$-mode  pulsations.  In agreement with previous  works, we found
that the  asymptotic period spacing  increases with a decrease  in the
stellar mass  and with  an increment in  the luminosity,  although the
dependence on  the stellar  mass is stronger.   This fact  renders the
value  of the  period  spacing observed  in  a given  star a  powerful
indicator of the total mass.

The study of the mode-trapping properties of our models has been other
relevant  aim  of  the  present  work. In  this  connection,  we  have
demonstrated that the mode-trapping features characterizing our PG1159
models are produced mostly by the shape of the O/C chemical profile at
the core, at least for the  range of periods observed in GW Vir stars.
On  their  hand,  the  outer  chemical interface  of  O/C/He  produces
negligible  mode-trapping  effects, except  for  the  regime of  short
periods in  the pulsation spectrum.   This conclusion is at  odds with
previous results reported by KB94, who realized that the mode-trapping
properties of their PG 1159 models were fixed essentially by the outer
O/C/He  transition region.   We have  found  that the  origin of  this
discrepancy rests on the differences  in the input physics employed in
build  up  the  background  stellar models.   Specifically,  the  main
difference  is  the  presence  of  a  much  less  pronounced  chemical
transition in the  outer parts of the C/O core of  the KB94 models, as
compared  with the rather  abrupt chemical  gradients at  $M_r \approx
0.4-0.6 M_*$ in our PG  1159 models. Other ingredient that contributes
to  the disagreement  between our  results and  those of  KB94  is the
thickness  of  the helium  envelope.   Indeed,  since  our models  are
characterized  by  thick helium  envelopes  ($M_{\rm  He} \sim  0.0052
M_{\odot}$), the mode trapping effects caused by the O/C/He transition
region are much more weaker than in the models of KB94.

We note that the structure of the core chemical profile of our PG 1159
models  is  the relic  of  convective  overshoot  episodes during  the
central helium  burning phase.  The  sensitivity of the  mode trapping
effects  to the  details  of  the core  chemical  structure rises  the
possibility of employing  pulsating PG 1159 and WD  stars to constrain
the  efficiency  of  extra  mixing  episodes  in  the  core  of  their
progenitors  (overshooting  and/or  semiconvection; see  Straniero  et
al.  2003).  This  appealing  issue  has  been  recently  explored  by
C\'orsico \& Althaus (2005).

To gain  additional insight  into the nature  of mode trapping  in our
models, we  have examined the  kinetic energy and weight  functions of
individual pulsation  modes. We have found that  modes showing primary
maxima in  the kinetic energy distribution are  associated with minima
in the period-spacing diagrams.   By consulting the eigenfunctions and
weight functions  of these  modes we found  that they  have relatively
high amplitude  at the high-density  environment of the  stellar core,
and very  low amplitudes in  the rest of  the star.  The  existence of
these ``core-trapped'' modes is encountered  also in the context of ZZ
Ceti pulsations, as reported by Althaus et al. (2003) and C\'orsico et
al.  (2005).

Finally,  we have made  some preliminar  seismic inferences  about the
internal  structure  and basic  parameters  of  the  GW Vir  stars  PG
1159-035, PG  2131+066, PG 1707+427, and  PG 0122+200. To  this end we
have  adopted  three different  approaches.   First,  we estimate  the
stellar  masses by comparing  the asymptotic  period spacing  with the
observed mean period spacing,  assuming that the effective temperature
of the target  star is that predicted by  spectroscopy.  In the second
method,  we repeated  the  above  procedure but  this  time using  the
average  of the  computed period  spacings and  comparing it  with the
observed mean period spacing.  The  third approach is a fitting method
in which we  search for an optimal stellar  model that best reproduces
the observed  periods. To this end  we employ a  quality function that
measures the distance between  the observed and the computed adiabatic
pulsation  periods within a  grid of  models with  different effective
temperatures and stellar masses.

The stellar masses  we obtained from these methods  are given in Table
1.  Note that, except for PG  0122+200, our values are consistent with
the  spectroscopic inferences,  particularly when  the average  of the
computed period  spacings are used,  although our masses  are slightly
larger.   For PG  0122+200 we  obtain, in  agreement with  the seismic
determination of  O'Brien (2000), a  large value of the  stellar mass.
Thus, according  the pulsation theory this  star seems to  be the most
massive among  the high-gravity GW  Vir pulsators. Note also  that the
stellar masses as determined by  employing the average of the computed
period  spacings are  appreciably lower  than the  values  inferred by
using  the asymptotic  period spacing.  We also  found that  our $M_*$
values are smaller than that  reported by other pulsation studies.  In
this connection, we note that, with  the exception of the work of KB94
for  PG 1159-035,  all these  works compare  the observed  mean period
spacing  with the asymptotic  predictions to  infer the  stellar mass.
Thus, not  surprisingly, their  values are very  close to  our results
obtained from  the asymptotic  period spacing, but  somewhat different
from  our values  deduced by  employing  the average  of the  computed
period spacings.

The results  from our period-fitting  procedure are shown in  Table 6.
Taking full advantage of the numerous pulsation periods observed in PG
1159-035,  we have  unambiguously obtained  a best-fit  model  with an
effective temperature  very close  to that predicted  by spectroscopy.
Our  asteroseismic mass  of $0.56  M_{\odot}$ is  consistent  with the
spectroscopic  calibration  ($M_*  =  0.54 M_{\odot}$)  and  with  the
preferred value  of Gautschy et  al. (2005) ($M_* =  0.55 M_{\odot}$),
but  considerably lower  than the  value  quoted by  KB94 ($M_*=  0.59
M_{\odot}$).   For the  remainder stars,  for which  we have  very few
observed  periods  available,  we  have also  obtained  representative
models  by employing  our period-fitting  procedure, but,  at variance
with the case  of PG 1159-035, for these stars we  have been forced to
consider  as  acceptable  solutions  only  models  with  an  effective
temperature laying within $1  \sigma$ of the spectroscopic value. This
stringent criteria  was necessary in  view of the numerous  and almost
equivalent minima  exhibited by the quality function  for these stars.
Note that the period matching are considerably poorer (Tables 3, 4 and
5) as compared with  the case of PG 1159-035 (Table  2).  We also have
estimated  the ``seismic''  distances to  the stars  by  employing the
luminosity of the  best-fit models. Our estimations (see  Table 6) are
consistent with other determinations.

Before closing the paper we  stress that, in our view, the pulsational
analysis  presented  here  constitutes  a substantial  improvement  as
compared  with  previous  studies.  However, we  believe  that  PG1159
evolutionary models with different  stellar masses based on a complete
description  of  the physical  processes  occurred  in  {\it all}  the
evolutionary stages  of progenitor stars would be  needed to reinforce
some  of our  results. The  development  of such  models is  certainly
critical  for  the assessment  of  the  internal chemical  composition
characterizing GW Vir stars, an issue  that is key as far as precision
asteroseismology is concerned.

Finally, the  evolutionary tracks  employed in this  investigation, as
well as tabulations of $\Pi_0$ in terms of $L_*$ and $T_{\rm eff}$ for
the complete  set of stellar masses  are freely available  at our URL:
{\tt http://www.fcaglp.unlp.edu.ar/evolgroup/}

\begin{acknowledgements}

We wish to thank  the  suggestions and comments of the 
anonymous referee that improved the original version of this work. 
This research was supported by the Instituto de Astrof\'{\i}sica  
La Plata (CONICET). 

\end{acknowledgements}

\appendix

\section{Pulsation equations and adiabatic quantities}

Our numerical pulsation code solves the fourth-order set of 
equations governing linear, 
nonradial, adiabatic stellar pulsations in the formulation 
given in Unno et al. (1989):

\begin{equation}
x \frac{dy_1}{dr}= \left(V_g-3\right)\ y_1 + \left[\frac{\ell(\ell+1)}
{C_1 \omega^2} - V_g \right]\ y_2 + V_g\ y_3,
\end{equation}

\begin{equation}
x \frac{dy_2}{dr}= \left( C_1 \omega^2 - A^* \right)\ y_1 + 
 \left( A^* - U + 1 \right)\ y_2 - A^*\ y_3,
\end{equation}

\begin{equation}
x \frac{dy_3}{dr}= \left( 1 - U \right)\ y_3 +\ y_4, 
\end{equation}

\begin{equation}
x \frac{dy_4}{dr}= U A^*\ y_1 + U V_g\ y_2 + 
\left[ \ell(\ell+1) - U V_g\right]\ y_3 - U\ y_4.
\end{equation}

\noindent The boundary conditions are, at the stellar 
centre ($x=0$):

\begin{equation}
y_{1}\ C_{1}\ \omega^2 - \ell\ y_{2} = 0,\\
\ell\ y_{3} - y_{4} = 0,
\end{equation}

\noindent and at the stellar surface ($x=1$):
		
\begin{equation}
y_{1} - y_{2} + y_{3} = 0,\\
(\ell+1)\ y_{3} + y_{4}= 0,
\end{equation} 

\noindent being the normalization $y_{1}= 1$ at $x=1$ ($x=r/R_*$). 
The dimensionless Dziembowski's variables (eigenvalue and eigenfunctions) 
are defined as

\begin{equation}
\omega_{k}^2= \frac{R_*^3}{G M_*} \sigma_{k}^2,
\end{equation}

\noindent and

\begin{equation}
y_1= \frac{\xi_r}{r},\\ \ y_2= \frac{1}{g r} 
\left(\frac{p'}{\rho}+\Phi'\right), 
\label{y1y2}
\end{equation}

\begin{equation}
y_3= \frac{1}{g r} \Phi',\\ \ y_4= \frac{1}{g} \frac{d \Phi'}{dr}.
\end{equation}

\noindent Here, $\xi_r$ is the radial Lagrangian 
displacement, and $p'$, $\Phi'$ are the Eulerian perturbation of 
the pressure and the gravitational potential, respectively. Pertinent 
dimensionless coefficients of the pulsation equations are:

\begin{equation}
V_g= \frac{g r}{c^2},\\ \ U=  \frac{4 \pi \rho r^3}{M_r},
\end{equation}

\begin{equation}
C_1= \left( \frac{r}{R} \right)^3 
\left( \frac{M_*}{M_r} \right), \\ \ A^*= \frac{r}{g} N^2, 
\end{equation}

\noindent where $c$ is the adiabatic local sound speed and $N$  
the Brunt-V\"ais\"al\"a frequency. The remainder symbols are 
self-explanatory. Once the eigenvalue and eigenfunctions are computed, 
the code proceeds to evaluate a number of important pulsation 
quantities, such as  the pulsation period,  $\Pi_{k}$,

\begin{equation}
\Pi_{k}= 2 \pi/\sigma_{k},
\end{equation}

\noindent the oscillation kinetic energy, $K_{k}$, 

\begin{eqnarray}
\lefteqn{ K_{k} = \frac{1}{2} (G M_* R_*^2) \omega_{k}^2} 
\nonumber \\
& & \times \int_{0}^{1} 
x^2 \rho \left[ x^2 y_1^2 + x^2 \frac{\ell (\ell +1)}{(C_1 
\omega_{k}^2)^2} y_2^2\right] dx,
\label{kinetic}
\end{eqnarray}

\noindent the weight function, $W_k$,  

\begin{eqnarray}
\lefteqn{W_k(x) = (4 \pi G R_*^2) \frac{x^2 \rho^2}{U}} \nonumber \\ 
& & \times \left[ A^* y_1^2 +  V_g \left( y_2 - y_3 \right)^2 
- \frac{1}{U} \{  
\ell \left(\ell + 1 \right) y_3 + y_4 \}^2 \right],
\label{wf}
\end{eqnarray}

\noindent the variational period, $\Pi_{k}^v$, 

\begin{equation}
\Pi_{k}^v= \sqrt{\frac{8 \pi^2}{G M_*}}\ 
\frac{K_{k}^{1/2}}{\omega_{k}} \left[ 
\int_0^1 W_k(x)\ x^2 dx \right]^{-1/2},
\end{equation}

\noindent and finally, the first-order rotation splitting 
coefficients, $C_{k}$,

\begin{equation}
C_{k}= \frac{(G M_* R_*^2)}{2 K_{k}}
\int_{0}^{1} 
\frac{x^2 \rho}{C_1} \left[ 2 x^2 y_1 y_2 + \frac{x^2}{C_1 \omega_{k}^2}
y_2^2\right] dx,
\end{equation}

\noindent The rotation splitting of the eigenfrequencies 
(assuming slow, rigid rotation) can be assesed by means of 

\begin{equation}
\sigma_{k,m}= \sigma_{k} + m(1-C_{k}) \Omega
\end{equation}

\noindent where $\Omega$ is the angular speed of rotation and $m$
the azimutal quantum number.

\end{document}